\documentclass{aastex63}

\newcommand\ia{\'{\i}}

\received{}
\revised{}
\accepted{\today}

\submitjournal{AJ}

\shorttitle{SOAR Speckle Orbits}
\shortauthors{Mendez et al.}

\begin{document}
    
\title{Orbits and masses of binaries from Speckle Interferometry at SOAR\footnote{Based on observations obtained at the Southern
    Astrophysical Research (SOAR) telescope, which is a joint project
    of the Minist\'{e}rio da Ci\^{e}ncia, Tecnologia, e
    Inova\c{c}\~{a}o (MCTI) da Rep\'{u}blica Federativa do Brasil, the
    U.S. National Optical Astronomy Observatory (NOAO), the University
    of North Carolina at Chapel Hill (UNC), and Michigan State
    University (MSU).}}

\correspondingauthor{Rene A. Mendez}
\email{rmendez@uchile.cl}

\author[0000-0003-1454-0596]{Rene A. Mendez}
\affiliation{Astronomy Department \\ Universidad de Chile \\ Casilla 36-D, Santiago, Chile}
  
\author{Rub\'en M. Claver\'ia}
\affiliation{Department of Engineering \\ University of Cambridge, UK}

\author{Edgardo Costa}
\affiliation{Astronomy Department \\ Universidad de Chile \\ Casilla 36-D, Santiago, Chile}

\begin{abstract}

We present results from Speckle inteferometric observations of fifteen visual binaries and one double-line spectroscopic binary, carried out with the HRCam Speckle camera of the SOAR 4.1 m telescope. These systems were observed as a part of an on-going survey to characterize the binary population in the solar vicinity, out to a distance of 250 parsec.

We obtained orbital elements and mass sums for our sample of visual binaries. The orbits were computed using a Markov Chain Monte Carlo algorithm that delivers maximum likelihood estimates of the parameters, as well as posterior probability density functions that allow us to evaluate their uncertainty. Their periods cover a range from 5~yr to more than 500~yr; and their spectral types go from early A to mid M - implying total system masses from slightly more than 4~$M_\odot$ down to 0.2~$M_\odot$. They are located at distances between approximately 12 and 200~pc, mostly at low Galactic latitude.

For the double-line spectroscopic binary YSC8 we present the first combined astrometric /radial velocity orbit resulting from a self-consistent fit, leading to individual component masses of $0.897 \pm 0.027~M_\odot$ and $0.857 \pm 0.026~M_\odot$; and an orbital parallax of $26.61 \pm 0.29$~mas, which compares very well with the Gaia DR2 trigonometric parallax ($26.55 \pm 0.27$~mas).

In combination with published photometry and trigonometric parallaxes, we place our objects on an H-R diagram and discuss their evolutionary status. We also present a thorough analysis of the precision and consistency of the photometry available for them.

\end{abstract}


\section{Introduction} \label{sec:intro}

The fundamental parameter determining the internal structure and evolution of stars of a given chemical composition is their initial mass (see e.g. \cite{Kahler1972}). The main relationship reflecting -and good for testing- the dependency of a star's properties on its mass is the observational mass-luminosity relation (MLR). Establishing the MLR is not straightforward because it involves determining precise masses and another elusive parameter: distance. To complicate things further, there seems to be an intrinsic dispersion in the MLR caused by differences in age and chemical composition from star to star (see e.g. \cite{Horchet2015,Horchet2019}).

Stellar masses can be obtained through the observation of binary systems. Considering that roughly half of solar-type stars in the Solar Neighborhood belong to binary systems \citep{DuchKra2013}, in principle it is possible to determine masses for a very large number of stars. On the other hand, the recent advent of the Gaia satellite \citep{Prustiet2016} has dramatically improved the precision of stellar distances within the Solar Neighborhood. Up to a distance of 250~pc, a parallax determined by Gaia has an uncertainty under 1\%, which by current standards has solved the distance dilemma in the MLR. In spite of this promising scenario, much remains to be done to increase the number of stars with well known masses and luminosities.

A starting point for systematic surveys to determine precise stellar masses are catalogues of visual binaries (confirmed or suspected), such as the Hipparcos catalogue \citep{Lindet1997}, or of spectroscopic binaries, such as the Geneva-Copenhagen spectroscopic survey \citep{Nordet2004}. To this end, these two samples have been exploited in this decade by Elliott Horch (Southern Connecticut State University) and collaborators, who have been doing Speckle imaging of binaries with  the WIYN 3.5m telescope at Kitt Peak (see e.g. \cite{Horchet2017}); and by Andrei Tokovinin (Cerro Tololo Inter-American Observatory) and collaborators who are using the HRCam Speckle camera of the SOAR 4.1 m telescope at Cerro Pachón, Chile (see e.g. \cite{Tokoet2014}).

In collaboration with the above research groups, since 2014 we have been carrying out an expanded Speckle campaign with HRCam@SOAR to determine orbits and masses of (mostly) southern systems included in the catalogues previously mentioned. In multiple observing runs, spanning 22 nights granted at SOAR (until March 2020), we have so far observed 1,719 distinct systems. Of these (which include suspected binaries from Hipparcos and tight spectroscopic binaries - see previous paragraph) we have been able to confirm, resolve, and measure 1,183 systems -many several times- (see e.g. \cite{Tokoet2015}, \cite{Gomezet2016}, \cite{Mendezetal2017}, \cite{Docoboet2019} and \cite{Tokoet2020}). Other notable results from our observations include the discovery of almost sixty inner or outer subsystems in previously known binaries, as well as two quadruple systems\citep{Tokoet2015,Tokoet2016}. More recently (in 2019), we have also started a program focused on very tight, short period systems present in our sample, with the ZORRO Speckle Camera of the Gemini South 8.1 m telescope at Cerro Pachón\footnote{See \url{https://www.gemini.edu/instrumentation/current-instruments/alopeke-zorro}}. Because of the small apparent angular separation between the components, these objects are difficult or impossible targets for a 4m class facility. 

Our expansion of the southern program (including a footprint to the north) will provide, in combination with the northern program, an all-sky, volume-limited, Speckle survey of binary systems. Surveying objects out to 250 pc from the Sun in both hemispheres, will permit to sample a larger volume in terms of Galactocentric distances and distances from the Galactic plane, allowing to encompass a broader range in metallicity and Galactic populations. When complete, our survey will permit sensitive tests of stellar evolution theory and add a significant number of new points to the MLR (for a recent effort to improve the MLR at masses less than 0.7~$M_\odot$ see \cite{Mannet2019}). With this we will be able to investigate effects such as metallicity and age on the MLR. In Section~\ref{sec:conc} we discuss how far we are from reaching this goal at the present time.

In this paper we report results for fifteen visual binaries, and one double-line spectroscopic binary, observed with HRCam@SOAR. In Section~\ref{sec:sample} we discuss the basic properties of our sample and in Section~\ref{sec:orbits} we present our results. In Section~\ref{sec:hr} we present an observational H-R diagram for our sample; in Section~\ref{sec:hr} we provide a list of comments for each object; and finally in Section~\ref{sec:conc} we give our conclusions. We note that in \cite{Mendezetal2017} we have discussed in full detail the procedures and methodology to determine the orbits; in this paper, we give more attention to the precision and consistency of the photometry available for our objects since, at present, this seems one of the major obstacles to make further progress in this field.

\section{Basic properties of the Binary Systems}\label{sec:sample}

As part of the standardized pipeline reductions of our SOAR/HRCam data, we constantly monitor the observed-computed ephemeris ([O-C]) for pairs with known orbital elements. The selection of targets presented in this paper is rather heterogeneous, and considers those orbital pairs for which their [O-C] in either angular separation ([O-C]$_\rho$) or
position angle ([O-C]$_\theta$) evaluated at the epoch of our SOAR
Speckle data was too large in comparison with the internal precision
of our data, indicative that their orbits should be revised or
improved with the addition of our new data points (in general, we also add binaries where first-time orbits could be computed using the SOAR
observations, albeit in this paper in particular there are none of these). Some objects of this initial list were later removed due to either a lack of sufficient historical data or due to the impossibility of improving on their orbits with the current orbital coverage.

Table~\ref{tab:photom1} gives basic properties available in the literature for the final sample analysed in this paper. The first three columns give the name in the Washington Double Star Catalogue (\cite{WDSCat2001}, hereafter WDS),
discoverer designation (WDS) and sequential number in the Hipparcos catalogue (HIP) - or an alternative name from the SIMBAD database \citep{SIMBAD}. The fourth column gives the apparent $V$ magnitude for the system listed in the SIMBAD database ($V_{\mbox{Simbad}}$). The fifth and sixth columns give the $V$ magnitude on the Hipparcos catalogue ($V_{\mbox{Hip}}$) and its source, respectively. The seventh and eighth columns list the values for the color ($(V-I)_{\mbox{Hip}}$) and source, respectively, also from the Hipparcos catalogue. The ninth and tenth columns give the $V$ magnitudes for the primary ($V_{\mbox{P}}$) and secondary ($V_{\mbox{S}}$) components,respectively, as listed in the WDS catalogue. As a sanity test, the integrated apparent magnitude for the system $V_{\mbox{Sys}}$ is given in the eleventh column\footnote{Computed as $V_{\mbox{Sys}} = -2.5 \times \log \left( 10^{-0.4 \cdot V_{\mbox{P}}} + 10^{-0.4 \cdot V_{\mbox{S}} } \right)$}. In the twelfth and thirteen columns we report our own measured magnitude differences in the Str\"omgren y filter ($\Delta y$) and in the Cousins I filter $\Delta I$ ($\equiv I_{\mbox{S}} -I_{\mbox{P}}$ between secondary and primary), respectively. Finally, in the last column we report the values for the spectral type and luminosity class for the primary (and secondary, after the $+$ sign, when available) from WDS and SIMBAD, respectively.

\floattable
\rotate
\begin{deluxetable}{cccccccccccccc}
\tablecaption{Object identification \label{tab:photom1}}
\tabletypesize{\scriptsize}
\tablecolumns{14}
\tablewidth{0pt}
\tablehead{
\colhead{WDS name} &
\colhead{Discoverer} &
\colhead{HIP number} & 
\colhead{$V_{\mbox{Simbad}}$\tablenotemark{a}} &
\colhead{$V_{\mbox{Hip}}$\tablenotemark{b}} &
\colhead{Source$_{V_{\mbox{Hip}}}$\tablenotemark{c}} &
\colhead{$(V-I)_{\mbox{Hip}}$\tablenotemark{b}} &
\colhead{Source$_{(V-I)_{\mbox{Hip}}}$\tablenotemark{d}} &
\colhead{$V_{\mbox{P}}$\tablenotemark{e}} &
\colhead{$V_{\mbox{S}}$\tablenotemark{e}} &
\colhead{$V_{\mbox{Sys}}$} &
\colhead{$\Delta y$\tablenotemark{f}} &
\colhead{$\Delta I$\tablenotemark{f}} &
\colhead{Sp Type} \\
\colhead{} &
\colhead{designation} &
\colhead{} & 
\colhead{} &
\colhead{} &
\colhead{} &
\colhead{} &
\colhead{} &
\colhead{} &
\colhead{} &
\colhead{} & 
\colhead{} &
\colhead{} &
\colhead{WDS/Simbad}
}
\startdata
 10043$-$2823 & I292 & 49336 & 7.29 & 7.27 & H & $0.54 \pm 0.01$ & H & 7.77 &
8.43 & 7.30 & $0.45 \pm 0.17$ & --- & F6V/F6V \\
10174$-$5354 & CVN16AaAb & TWA22\tablenotemark{g} & $13.96 \pm 0.01$ & --- &
--- & --- & --- & 14.51\tablenotemark{h} &  14.87\tablenotemark{h} & 13.8 & $0.58 \pm 0.10$ & --- & M6+M6/M5 \\
11125$-$1830 & BU220 & 54742 & $6.105 \pm 0.010$ & 6.11 & H & $0.01 \pm 0.01$
& L & 6.24 & 8.34 & 6.09 & $1.10 \pm 0.10$ & 0.85 & A1V/A0V \\
13145$-$2417 & FIN297AB & 64603 & $6.68 \pm 0.01$ & 6.66 & H & $0.30 \pm 0.01$
& L & 7.48 & 7.51 & 6.74 & $0.35 \pm 0.10$ & --- & F0V/F0V \\
13574$-$6229 & FIN370 & 68170 & $6.65 \pm 0.01$ & 6.65 & G & 0.82 &
H & 7.37 & 7.76 & 6.80 & $0.31 \pm 0.11$ & --- & G3III-IV/G3III-IV \\
15006$+$0836 & YSC8 & 73449 & --- & 7.26 & G & 0.75 & H & 7.5 & 9.4 & 7.33 & $0.13 \pm 0.12$ & $0.27 \pm 0.06$ & G0V/G0V  \\
15160$-$0454 & STF3091AB & 74703 & 7.29 & 7.21 & G & 0.58 & H & 7.74 & 8.48 &
7.30 & 0.50 & 0.60 & F8V/F7V \\
15420$+$0027 & A2176 & 76892 & 7.24 & 7.23 & H & $0.27 \pm 0.02$ & H & 8.0 &
8.0 & 7.25 & $0.63 \pm 0.19$ & --- & A0/A1IV \\
17005$+$0635 & CHR59 & 83223 & $6.566 \pm 0.010$ & 6.57 & H & $0.26 \pm 0.01$
& H & 6.6 & 9.9 & 6.55 & $3.10 \pm 0.17$ & $2.28 \pm 0.21$ & A7V/A7V \\
17077$+$0722 & YSC62 & GJ1210\tablenotemark{g} & --- & --- & --- & --- & --- & 14.4 & 15.4 & 14.04 & --- & $0.13 \pm 0.15$ & M5/M5V \\
17155$+$1052 & HDS2440 & 84417 & --- & 8.67 & H & $0.78 \pm 0.01$ & L & 9.24 &
10.04 & 8.82 & --- & $0.17  \pm 0.15$ & G5/G5 \\
17181$-$3810 & SEE324 & 84634 & $6.92 \pm 0.01$ & 6.92 & H & $0.05 \pm 0.01$ &
L & 7.51 & 7.91 & 6.94 & $0.53 \pm 0.08$ & 0.40 & A0IV/A0IV \\
17283$-$2058 & A2244AB & 85491 & 7.94 & 7.93 & H & $0.55 \pm 0.01$ & L & 8.68
& 8.9 & 8.03 & 0.80 & 0.25 & F5V/F5V \\
17571$+$0004 & STF2244, & 87875 & $5.966 \pm 0.010$ & 5.95 & G & $0.12 \pm 0.01$ & H & 6.56 & 6.89 & 5.96 & 0.50 & 0.30 & A3V/A3V \\
19190$-$3317 & I253AB & 94926 & 6.96 & 6.94 & G & 0.65 & H & 7.25 &
8.77 & 7.01 & 0.20 & 0.25 & G1V/G1V \\
19471$-$1953 & BU146 & 97348 & $8.25 \pm 0.01$ & 8.17 & H & $0.66 \pm 0.01$ &
L & 8.61 & 9.52 & 8.22 & 1.25 & 1.00 & G1-2V/G1-2V \\
\enddata
\tablenotetext{a}{From SIMBAD.}
\tablenotetext{b}{From Hipparcos catalogue.}
\tablenotetext{c}{G=ground-based, H=HIP.}
\tablenotetext{d}{'H' when $V-I$ derived from measurements in other
  bands/photoelectric systems than the Cousins' system; 'L' when $V-I$ derived from
  Hipparcos and Star Mapper photometry.}
\tablenotetext{e}{From WDS.}
\tablenotetext{f}{From our own measurements in the Str\"omgren y-band or in
  the Cousins' $I$-band respectively. When one more than one measurement, it is the average, excluding uncertain (:) values. When more than three measurements, we quote the standard deviation on the mean.}
\tablenotetext{g}{Not in Hipparcos, alternative name.}
\tablenotetext{h}{Not from WDS, but calculated in Section~\ref{sec:comp} for a $V_{\mbox{Sys}} = 13.92$.}
\tablenotetext{i}{The Tycho catalogue reports for this object $V_T=7.167$ and $B_T=7.488$. Using the color transformation in \citet{Mil2005} for $B_T-V_T = 0.321 \sim B-V$, it implies $V \sim 7.13$  (see his Figures~1 and~2), in agreement with the value reported by SIMBAD.}
\end{deluxetable}

\subsection{Photometry}\label{sec:photom}

The consistency of the photometry is important when addressing the compatibility of the dynamical and trigonometric parallaxes, or when comparing the astrometric mass sum to the dynamical masses (see Section~\ref{sec:massum}). Precise photometry is also needed to place the individual components in an HR diagram (Section~\ref{sec:hr}). While there is an overall good agreement between the SIMBAD, Hipparcos, and the combined ($V_{\mbox{Sys}}$) magnitudes, the quality of the photometry presented in Table~\ref{tab:photom1} is somewhat variable, as can be readily seen by comparing the fourth, fifth, and eleventh columns of that table.

In order to increase our comparison basis, we have searched for available photometry of our targets in more recent all-sky photometric surveys for bright stars; in particular "The All Sky Automated Survey" (ASAS\footnote{\url{http://www.astrouw.edu.pl/asas/?page=main}}
\cite{Pojmanski1997}); the "All-Sky Automated Survey for Supernovae"
(ASAS-SN\footnote{\url{http://www.astronomy.ohio-state.edu/asassn/index.shtml}},
\cite{Kochaneket2017,Jayasingheet19}); and "The AAVSO Photometric All-Sky Survey" (APASS\footnote{\url{https://www.aavso.org/apass}}, \cite{Hendenet2009}, data release 10, November 2018).
All three catalogues report $V$-band magnitudes. APASS incudes in addition Sloan i'-band photometry, which unfortunately cannot be compared directly with $I$ band values from Hipparcos. To have an extra comparison source in the I filter, we used the "All-sky spectrally matched Tycho-2 stars" available at the CDS\footnote{VizieR Catalogue VI/135}. This catalogue presents synthetic photometry in various bands, including $V$ and $I$, from an spectral energy distribution (SED) fit to 2.4 million stars in the Tycho-2 catalogue by \cite{PicklesDepagne2010} (PD2010 hereafter).

In Table~\ref{tab:photom2} we present the photometry obtained from the above catalogs, together with their quoted uncertainties. In some cases these latter are our own estimations, from the published light curves (see Figures~\ref{fig:photom2} and~\ref{fig:photom3} in Appendix~\ref{A1}).

While the above surveys measure and report everything they detect, their photometry is not reliable at the bright end. Based on the description of the different surveys, the reliability limit for ASAS, ASAS-SN, and APASS is 8, 10 and 7th mag respectively. Therefore, in Table~\ref{tab:photom2} we indicate with a colon (:) dubious values (up to 1 mag brighter than the bright mag limit per each survey), and with a double colon (::) even brighter objects whose measurements should be considered as very uncertain.

\floattable
\begin{deluxetable}{cccccccc}
\tablecaption{Additional photometry \label{tab:photom2}}
\tabletypesize{\scriptsize}
\tablecolumns{8}
\tablewidth{0pt}
\tablehead{
\colhead{WDS name} &
\colhead{Discoverer} &
\colhead{$V_{\mbox{ASAS}}$} &
\colhead{$V_{\mbox{ASAS-SN}}$} &
\colhead{$V_{\mbox{APASS}}$} &
\colhead{$i'_{\mbox{APASS}}$} &
\colhead{$V_{\mbox{PD2010}}$} &
\colhead{$I_{\mbox{PD2010}}$} \\
\colhead{} &
\colhead{designation} &
\colhead{} &
\colhead{} &
\colhead{} &
\colhead{} &
\colhead{} &
\colhead{}
}
\startdata
10043$-$2823 & I292 & $7.262 \pm 0.058$ : & --- & $7.31 \pm 0.22$ & $6.25 \pm 0.69$ :
&
7.286 & 6.71 \\ 
10174$-$5354 & CVN16AaAb & $14.16 \pm 0.20$\tablenotemark{a} & $13.914 \pm
0.038$\tablenotemark{a} & 13.969 & $11.388 \pm 0.052$ & 10.091 & 9.91 \\
11125$-$1830 & BU220 & $6.42 \pm 0.36$ :: & --- & --- & --- & 6.062 & 6.1 \\
13145$-$2417 & FIN297AB & $7.200 \pm 0.093$ : & --- & --- & --- & 6.703 & 6.37 \\
13574$-$6229 & FIN370 & $6.660 \pm 0.028$ :: & $7.43 \pm 0.11$ :: & --- & --- & 6.647 &
5.77 \\
15006$+$0836 & YSC8 & $7.237 \pm 0.027$ : & --- & $7.50 \pm 0.11$ & $6.887 \pm 0.058$ & 7.29 & 6.55 \\
15160$-$0454 & STF3091AB & $7.232 \pm 0.026$ : & $7.99 \pm 0.19$ :: & $7.35 \pm 0.11$ &
$6.90 \pm 0.11$ & 7.316 & 6.71 \\ 
15420$+$0027 & A2176& $7.213 \pm 0.025$ : & --- & $7.32 \pm 0.35$ & $7.32 \pm 0.31$ &
7.261 & 7.04 \\
17005$+$0635 & CHR59 & $6.66 \pm 0.17$ :: & --- & --- & --- & 6.587 & 6.37 \\
17077$+$0722 & YSC62 & $14.00 \pm 0.24$\tablenotemark{a} & --- & $14.161 \pm 0.038$ & $11.739 \pm 0.067$ &
--- & --- \\
17155$+$1052 & HDS2440 & $8.672 \pm 0.030$\tablenotemark{a} & $9.14 \pm 0.12$ : & $8.577 \pm 0.002$ &
$8.236 \pm 0.003$ & 8.65 & 7.91 \\
17181$-$3810 & SEE324 & $6.948 \pm 0.036$ :: & $9.64 \pm 0.53$ : & $6.969 \pm 0.001$ :
& $7.386 \pm 0.002$ & 6.866 & 6.88 \\
17283$-$2058 & A2244AB & $7.948 \pm 0.029$ : & $8.41 \pm 0.49$ :: & $8.875 \pm 0.002$ &
$7.747 \pm 0.002$ & 7.923 & 7.35 \\
17571$+$0004 & STF2244 & $6.30 \pm 0.28$ :: & --- & $5.978 \pm 0.010$ :: & $6.182 \pm 0.001$
& 6.006 & 5.90 \\
19190$-$3317 & I253AB & $6.908 \pm 0.024$ :: & --- & $7.050 \pm 0.001$ & $6.676 \pm
0.002$ : & 6.964 & 6.27 \\ 
19471$-$1953 & BU146 & $8.189 \pm 0.022$\tablenotemark{a} & ---& $8.185 \pm 0.033$ & $8.06 \pm 0.16$ &
8.237 & 7.51 \\
\enddata
\tablenotetext{a}{From our calculation, based on the light curves (see Figures~\ref{fig:photom2} and~\ref{fig:photom3} in Appendix~\ref{A1}).}
\end{deluxetable}

In Figure~\ref{fig:photom1} we show a comparison of the values presented in Tables~\ref{tab:photom1} and~\ref{tab:photom2} for the $V$ (upper panel) and $I$ (lower panel) filters. We chose to plot the Hipparcos $V$ and $I$ mag in the abscissa because it is the largest and more homogeneous dataset for our sample of targets. Two of our objects lack Hipparcos photometry, and are hence not included in these plots; namely CVN16AaAb 
and YSC62, they are of course too faint to be on the Hipparcos catalogue.

As can be seen in this figure (top panel), in the $V$ band there is a good correspondence between the photometry from Hipparcos and that from SIMBAD (which comes from different heterogeneous sources), and also with the combined photometry $V_{\mbox{Sys}}$ from WDS. The fit of $V_{\mbox{Sim}} {\it vs.} V_{\mbox{Hip}}$ has an rms residual of 0.026~mag, while that of $V_{\mbox{Sys}}$ {\it vs.} $V_{\mbox{Hip}}$ is 0.056~mag. We will adopt these values as an estimate of the uncertainty of the photometry in Section~\ref{sec:hr} (see also Figure~\ref{fig:hrdiag}). 

Consistent with what is mentioned above, at the bright end ($V < 6.8$) the ASAS and APASS magnitudes exhibit large photometric errors and scatter. In the case of ASAS-SN this problem extends down to the faintest data plotted ($V \sim 8$). ASAS-SN $V$ for object SEE324 is an extreme outlier: at $V_{\mbox{Hip}}=6.92$ and $V_{\mbox{ASAS-SN}}=9.64 \pm 53$, this object falls outside this plot, while the measurements from all the other surveys cluster around the one-to-one relationship (we have positively confirmed the identification of this target).

In the range $V > 7.0$ both ASAS and APASS exhibit good consistency, within the errors, between each other and also with Hipparcos and SIMBAD. The sole exception is A2244AB at $V_{\mbox{Hip}}=7.93$ for which $V_{\mbox{APASS}}=8.875$. The large rms value for $V_{\mbox{ASAS-SN}}$ for this object (0.49~mag, see Table~\ref{tab:photom2}), might be a hint that it is a variable source - if the scatter is due to intrinsic brightness variations and not due to saturation effects (note however the small rms from ASAS and APASS)\footnote{Note also the consistency between the SIMBAD, Hipparcos and the System's magnitude from the WDS in Table~\ref{tab:photom1}.}. It is not listed as a variable object by ASAS-SN, but, as discussed in Section~\ref{sec:hr}, the APASS value for this object renders better consistency between the dynamical and astrometric parallax and masses.

For the $I$ band (lower panel on Figure~\ref{fig:photom1}), the comparison is restricted only to the PD2010 SED fitted photometry and the Sloan i' filter measurements from the APASS survey. The high-residual point shown in the plot corresponds to I292, but this is not inconsistent given the very large error reported in Table~\ref{tab:photom2}, albeit its APASS $V$-band magnitude is consistent with that from Hipparcos. We have computed the difference $i' - I_{\mbox{PD2010}}= 0.35 \pm 0.12$ and $i' - I_{\mbox{Hip}}= 0.388 \pm 0.084 $ ($I_{\mbox{Hip}}= V_{\mbox{Hip}} - (V-I)_{\mbox{Hip}}$) excluding two outliers (I292 and CVN16AaAb - the latter not shown on Figure~\ref{fig:photom1} since being too faint it does not have an I-band magnitude from Hipparcos\footnote{Note that the V-band synthetic PD2010 photometry for CVN16AaAb is also an outlier, see Table~\ref{tab:photom2} and Figure~\ref{fig:photom2} in Appendix~\ref{A1}.}). The mean offset of 0.38 mag is shown by the thin dashed line in the figure, which indicates that the APASS $I$-band photometry is commensurable to that derived by Hipparcos and PD2010, after applying this offset. In Appendix~\ref{A1} we present a more detailed comparison of the consistency of the published photometry for individual targets from these surveys.

\begin{figure}[ht!]
\plotone{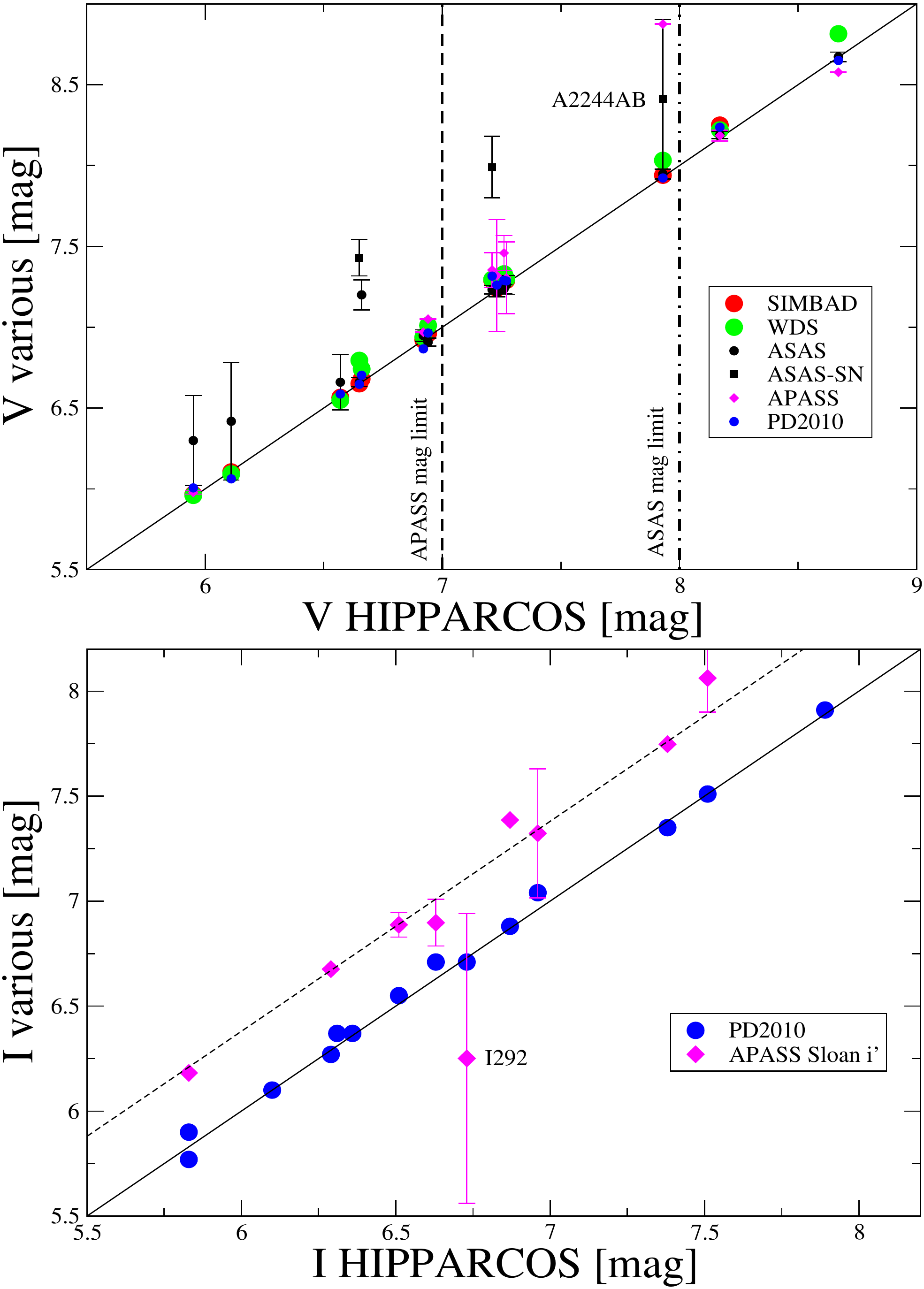}
\caption{Comparison of Hipparcos $V$ (top panel) and $I$ magnitudes (bottom panel) with other photometric data included in Tables~\ref{tab:photom1} and~\ref{tab:photom2}. The blue circles in both panels (labelled PD2010) are from SED fittings in \cite{PicklesDepagne2010}, as explained in the text, the magenta diamonds are for APASS. The dotted vertical lines show the bright limits for reliable photometry of the APAS ($V=7$) and ASAS ($V=8$) surveys, as indicated. In the whole magnitude range covered by these figures, {\it all} ASAS-SN photometry is unreliable (note their large declared error bars), but it was included for completeness. In both panels, the diagonal line depicts a one-to-one relationship. In the lower panel, the diagonal dashed line shows the 0.38 mag offset corresponding of the Sloan~$i'$ photometry from APASS. Highly discrepant points in these plots are discussed in the text.\label{fig:photom1}}
\end{figure}

\subsection{Individual component magnitudes}\label{sec:comp}

In what follows we will use the systems photometry $V_{\mbox{Sys}}$ from WDS -and the corresponding individual component magnitudes- listed in Table~\ref{tab:photom1}, beause these data are in overall agreement with those from Hipparcos. The comparisons made in the previous section, show that there is good consistency between the photometry from the different catalogues available, which deems reasonable to adopt data from these sources in the case of objects with missing photometric entries in Table~\ref{tab:photom1}. 

One object, CVN16AaAb does not have individual component magnitudes listed in WDS nor Hipparcos photometry, but it does have a combined magnitude, $V_{\mbox{Sys}} =13.8$. This combined magnitude is however somewhat inconsistent with the value listed in SIMBAD, which happens to be consistent with those from ASAS, ASAS-SN and APASS (see Table~\ref{tab:photom2} and Figure~\ref{fig:photom1}). Therefore, for this object we have adopted instead  $V_{\mbox{Sys}} =13.924 \pm 0.037$ which results from a (weighted) mean from ASAS, ASAS-SN and APASS (for APASS we have adopted the same error given in ASAS, because no error is given in the catalogue). Since we do not have a $\Delta V$ for this target, in order to compute the individual component magnitudes, we need to use our own measured magnitude difference. It is customary to adopt\footnote{Andrei Tokovinin, personal communication} $\Delta V = \Delta y$. Using 206 binary systems \cite{TokMasHar2010} have determined that $<\Delta y - \Delta V> = +0.23 \pm 0.30$ (see their Table 3). In order to verify this, from Table~\ref{tab:photom1}, excluding three outliers (BU220, I253AB, and YSC8), we compute $<\Delta V - \Delta y> = -0.22 \pm 0.30$, in agreement with Tokovini's result. Using this offset, plus our measured $\Delta y$ reported for CVN16AaAb in the twelfth column of Table~\ref{tab:photom1}, we adopt $\Delta V = 0.36$. Finally, the individual component magnitudes were computed as (primary) $V_{\mbox{P}} = V_{\mbox{Sys}} + 2.5 \times \log \left( 1.0 + 10^{-0.4 \cdot \Delta V} \right) = 14.51$ and (secondary) $V_{\mbox{S}} = V_{\mbox{P}}+ \Delta V = 14.87$, these are the values shown in that Table (for a further discussion of this object, please see Section~\ref{sec:hr}).

In the $I$-band, a fair fraction of our targets have a measured value of $\Delta I$ (column thirteen in Table~\ref{tab:photom1}). For these, the individual component magnitudes are computed in a similar manner as as was done for the $V$-band, except that the (combined) magnitude for each system was computed from $I_{\mbox{Sys}}  = V_{\mbox{Sys}} - (V-I)_{\mbox{Hip}}$. Note that to compute $I_{\mbox{Sys}}$ we used $V_{\mbox{Sys}}$ rather than $V_{\mbox{Hip}}$ so that the derived pairs $(V,I)_{\mbox{P,S}}$ are self-consistent (albeit, in general, as noted in Section~\ref{sec:photom}, there is good agreement between $V_{\mbox{Sys}}$ and $V_{\mbox{Hip}}$). Unfortunately, we have not measured  $\Delta I$ on SOAR for targets
I292, CVN16AaAb, FIN297AB, FIN370, and A2176, and therefore we can not compute the individual component magnitudes in this band for these objects. For target YSC62 we do not have $(V-I)_{\mbox{Hip}}$, but we have measured its $\Delta I$ on SOAR. We can estimate its $I_{\mbox{APASS}} \equiv i'_{\mbox{APASS}}-0.38= 11.36$ (see Section~\ref{sec:photom} and lower panel on Figure~\ref{fig:photom1}), and its color is then taken as $V_{\mbox{APASS}}- I_{\mbox{APASS}}= 2.80$. Additionally, YSC62 does not have a $V$-band magnitude from SIMBAD nor HIPPARCOS, but it does have $V_{\mbox{Sys}}$ from WDS, and we can thus proceed to compute the individual component magnitudes in the $I$-band in the same way as explained previously.


Our magnitude differences were computed using the method extensively described in \cite{TokMasHar2010} (see their Section 2.6)), which involves measuring the ratio of the peaks in the auto-correlation function derived from the power spectrum by Fourier transform of the spatial images. It is shown that the accuracy and precision of the magnitude difference depends on the signal-to-noise of the images, the magnitude difference itself, and the angular separation of the binary, leading to intrinsic random errors of the derived photometry of around 0.2 mag rms, and a bias (overestimation) in $\Delta m$ (see our comparison of $\Delta V$ to $\Delta y$ mentioned in the previous paragraphs\footnote{Note that $\Delta y$  are our measured values, whereas $\Delta V$ is taken from the WDS catalogue (usually taken from the resolved Hipparcos photometry) which, presumably, is free from this overestimation}). A partial solution to the bias has been proposed by \cite{Tokoet2015} (their  Section~2.3), but it is applicable only to wider pairs. Alternative promissory methods, which can in principle yield up to 0.02 mag of accuracy have been proposed to measure magnitude differences using speckle data, e.g, those in \cite{Baleet2002} and \cite{Pluz2005} which however requires further development in order to circumvent some of the instabilities in the estimation method, as reported by \cite{Pluz2005}, and a more systematic comparison using larger binary star datasets of speckle observations to validate these methods.

A final relevant comment regarding the overall uncertainty of our SOAR/HRCam photometry: all our inter-comparisons in the (combined) $V$ and $I$-bands imply an rms for the magnitudes and colors of about 0.1 mag or slightly less. It is more difficult to ascertain the uncertainty of our $\Delta y$ and $\Delta I$ values, because it depends on a number of factors such as the angular separation between the components, the quality of the sky when the measurement was performed, the brightness of the primary, etc. From repeated measurements on different nights for several of our objects (see Table~\ref{tab:photom1}), we estimate an average uncertainty of $\Delta y, \Delta I \sim 0.13$~mag, which we take as the typical error for our sample of binaries. This, typically poor, precision of the individual component magnitudes further limits a finer analysis and interpretation of the location of the components in an H-R diagram (see Figure~\ref{fig:hrdiag}), as explained in Section~\ref{sec:hr}. On the other hand, current precision of the astrometric interferometric measurements, in conjunction with the trigonometric parallaxes provided by the Gaia satellite \citep{Lurietal2018}, have dramatically changed the traditional limitations of visual binary work (see Section~\ref{sec:conc} for further discussion of this point).

\subsection{SOAR HRCam astrometry}\label{sec:astro}

The orbits presented in this paper are based on recent measurements made with the HRCam speckle camera mounted on the SOAR 4.1 m telescope, in the context of the program described by \cite{Mendezetal2017}. HRCam is described in \cite{TokCan2008}\footnote{For up-to-date details of the instrument see \url{http://www.ctio.noao.edu/~atokovin/speckle/}}.
Our observations (all of which are regularly ingested into the USNO Fourth Catalog of Interferometric Measurements of Binary Stars\footnote{The latest version, called int4, can be consulted here: \url{https://www.usno.navy.mil/USNO/astrometry/optical-IR-prod/wds/int4/fourth-catalog-of-interferometric-measurements-of-binary-stars}}, have been supplemented with historical astrometric measurements and computed orbital parameters (if available), compiled as part of the WDS effort, which were kindly provided to us by Dr. Brian Mason from the US Naval Observatory\footnote{In the site 
\url{http://www.das.uchile.cl/~rmendez/0001_Research/2_Visual_Binaries/Mendez_Claveria_Costa_AJ_2020/} we make available our input files, indicating the adopted uncertainty and quadrant flips (if any) for each data entry (see Section~\ref{sec:orbits}), and the origin of the measurements in the last column, following the nomenclature in int4.}.

Details of the data processing and calibration of HRCam are explained in \cite{TokMasHar2010}. Regarding the precision and accuracy of our astrometric data, the reader is referred to the publications from which these data were taken, which consists of the series of publications starting with \cite{TokMasHar2010} all the way to our 2019 data, reported on \cite{Tokoet2020}, where these issues are extensively discussed. In a nutshell, HRCam routinely delivers a precision of 1-3~mas in angular separation for objects brighter than $V \sim 12$. On every observing night we include "calibration binaries", these are binaries with very well known orbits (grade 5 on the USNO orbit catalogue), from which we calibrate our measurements, leading to systematic errors less than 0.1~deg in position angle, and better than 0.2\% in scale, i.e., smaller than our internal precision. The final precision of our measurements depends on a number of factors, but in this paper we adopt a uniform uncertainty of 3 mas, which is representative of all our measurements. As for the uncertainty (or equivalent weight) of the historical data (necessary for the orbit calculations, see Section~\ref{sec:orbits}), we adopted the value indicated in the WDS entries when available, or estimated the errors depending on the observation method (e.g., interferometric {\it vs.} digital or photographic or micrometer measurements). As emphasized by \cite{Mendezetal2017}, one should bear in mind that the assignment of weights to each observational point (a process that is somewhat "subjective", specially for older data) plays a an important role in the orbital solution, being sometimes responsible for slightly different orbital solutions from different authors, using the same basic astrometric dataset. 

\section{Orbits}\label{sec:orbits}

\subsection{Orbital elements for the visual binaries}\label{sec:orbel}

For our orbital calculations we have used a Bayesian MCMC orbital code with dimensionality reduction, whose implementation is described in detail in \cite{Mendezetal2017} and \cite{Clavet2019}. The main motivation behind this approach is to exploit features that are inherent to these methods, namely (i) to provide realistic confidence limits to the derived orbital elements, and (ii) to generate posterior probability density functions (PDF hereafter) for each orbital element, as well as for the derived masses. Since the convergence of MCMC methods tends to be faster (and more reliable) if one initializes the algorithm from a point close to the global minimum, we have used as starting orbital parameters to "feed" our MCMC code those derived from the versatile IDL-driven interactive ORBIT code developed by \citet{Toko1992}\footnote{The code and user manual can be downloaded from \url{http://www.ctio.noao.edu/~atokovin/orbit/index.html}}, which employs a parametric $\chi^2$ minimization approach.

The results of our MCMC code as applied to the fifteen visual binaries of the sample presented in Section~\ref{sec:sample} are shown in Table~\ref{tab:orbel1}. For each object, two sets of numbers for the orbital elements are provided: The upper row represents the configuration with the smallest mean square sum of the O-C overall residuals,
while the lower row shows the median derived from the posterior PDF of the MCMC simulations, as well as the upper (third) quartile ($Q75$) and lower (first) quartile ($Q25$) of the distribution in the form of a superscript and subscript respectively\footnote{We choose to report the quartiles instead of the classical $\sigma$, because they are more meaningful when the PDFs exhibit long tails, i.e., on not well determined orbits (see \cite{Mendezetal2017}), while for a Gaussian function they are equivalent: $\sigma = (Q75-Q25)/1.349$.}.

\begin{figure}[ht]
\begin{minipage}[b]{0.5\linewidth}
\centering
\includegraphics[width=.99\textwidth]{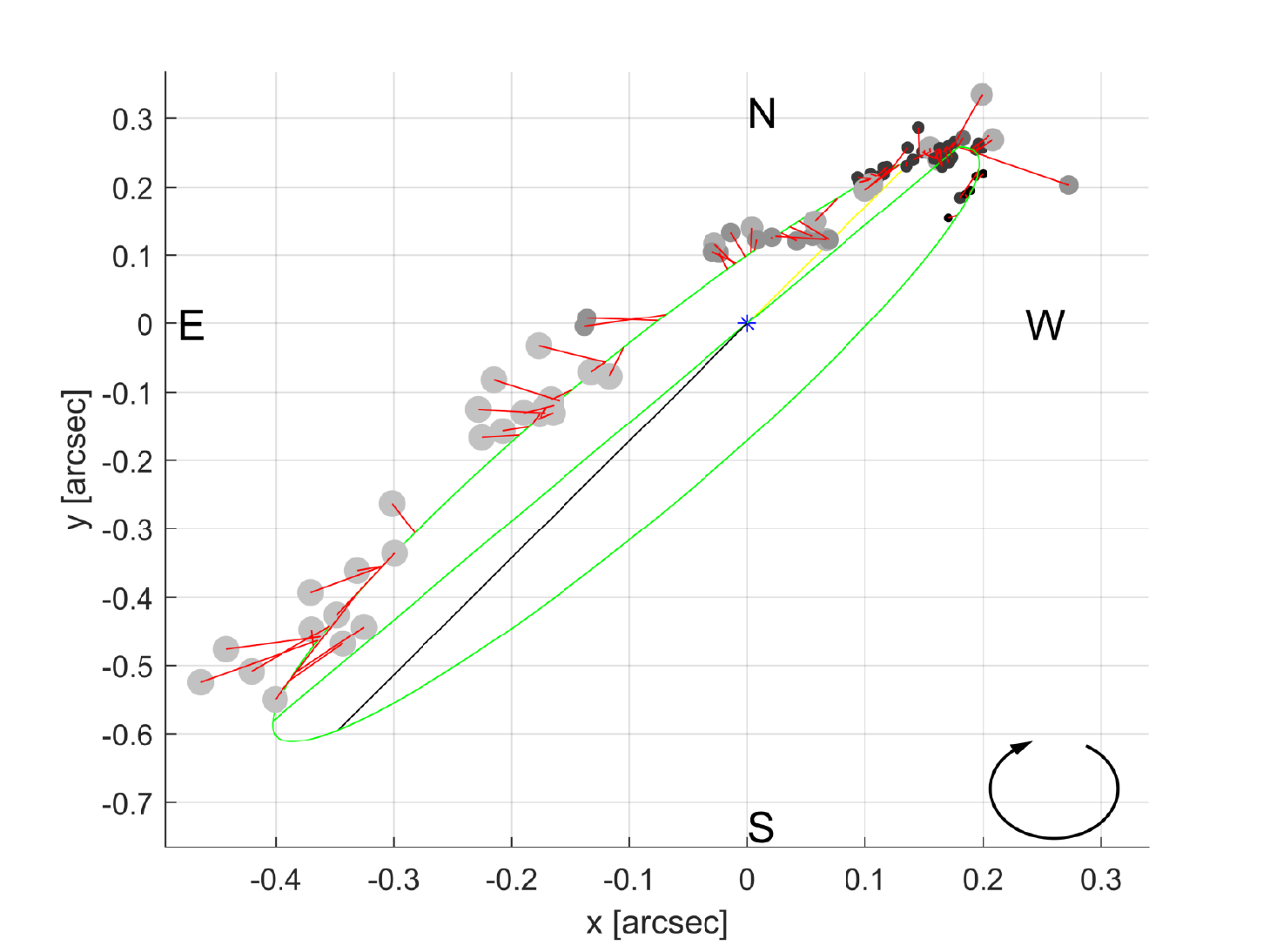}\\
\includegraphics[width=.99\textwidth]{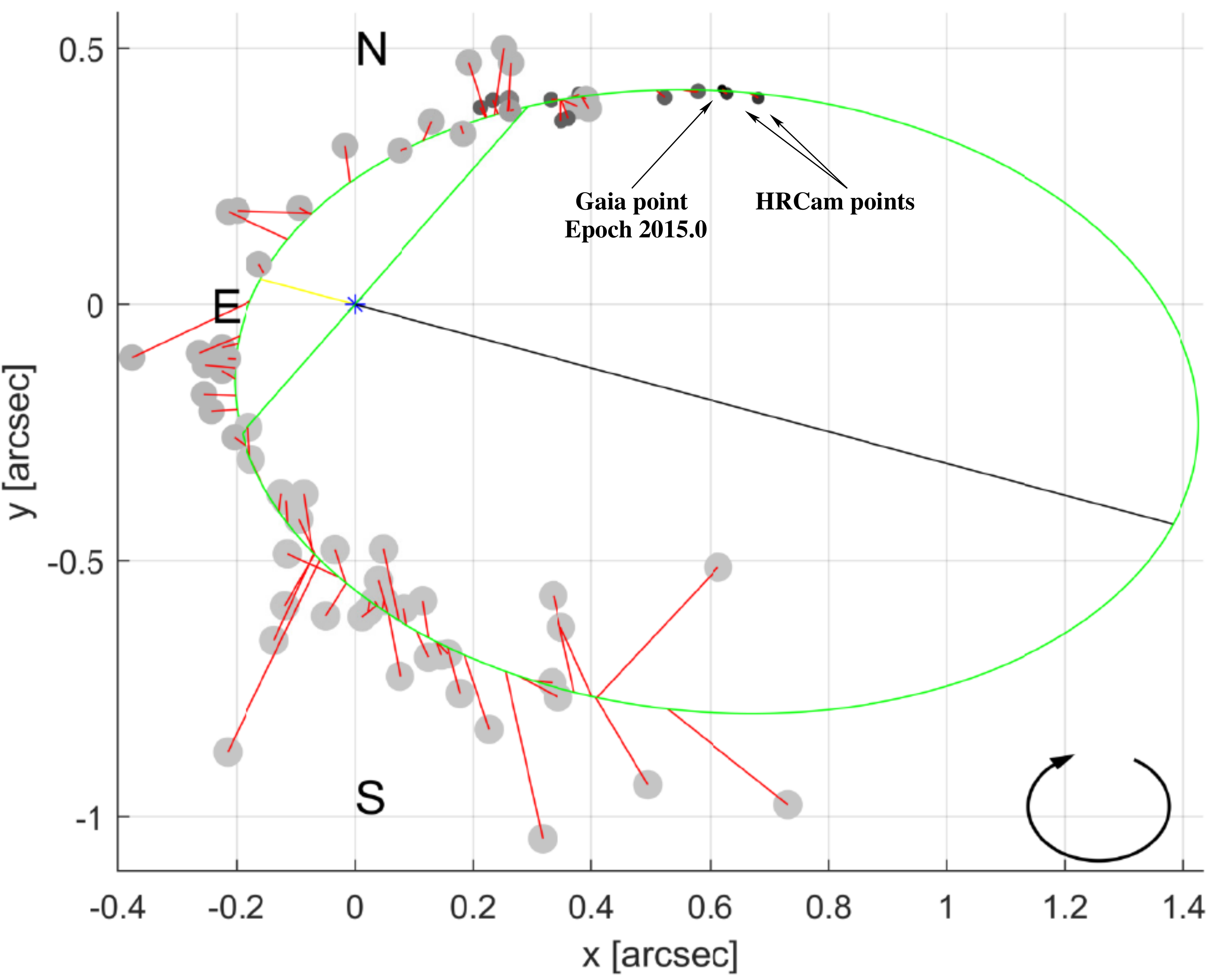}\\
\includegraphics[width=.99\textwidth]{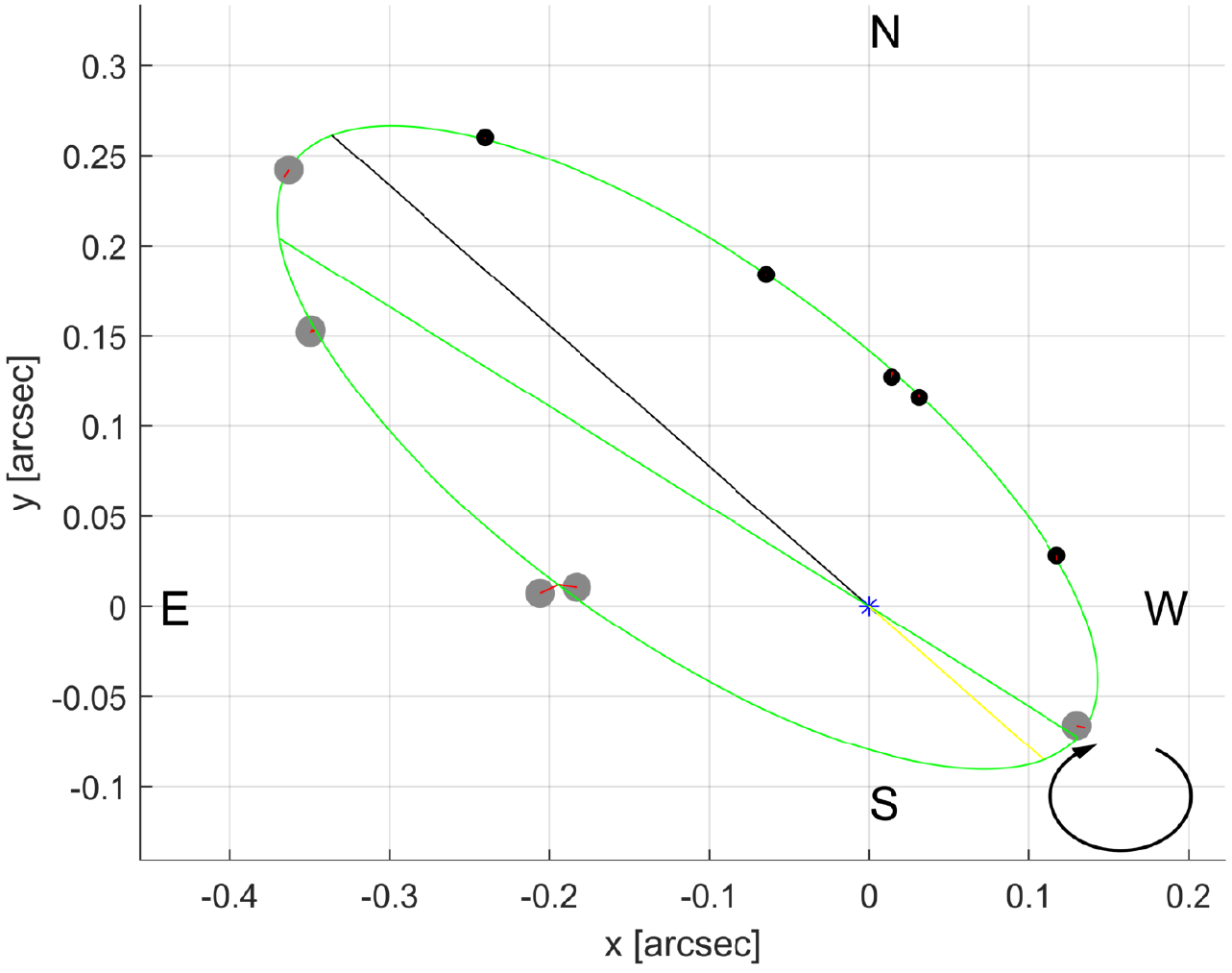}
\end{minipage}
\begin{minipage}[b]{0.5\linewidth}
\centering
\includegraphics[width=.99\textwidth]{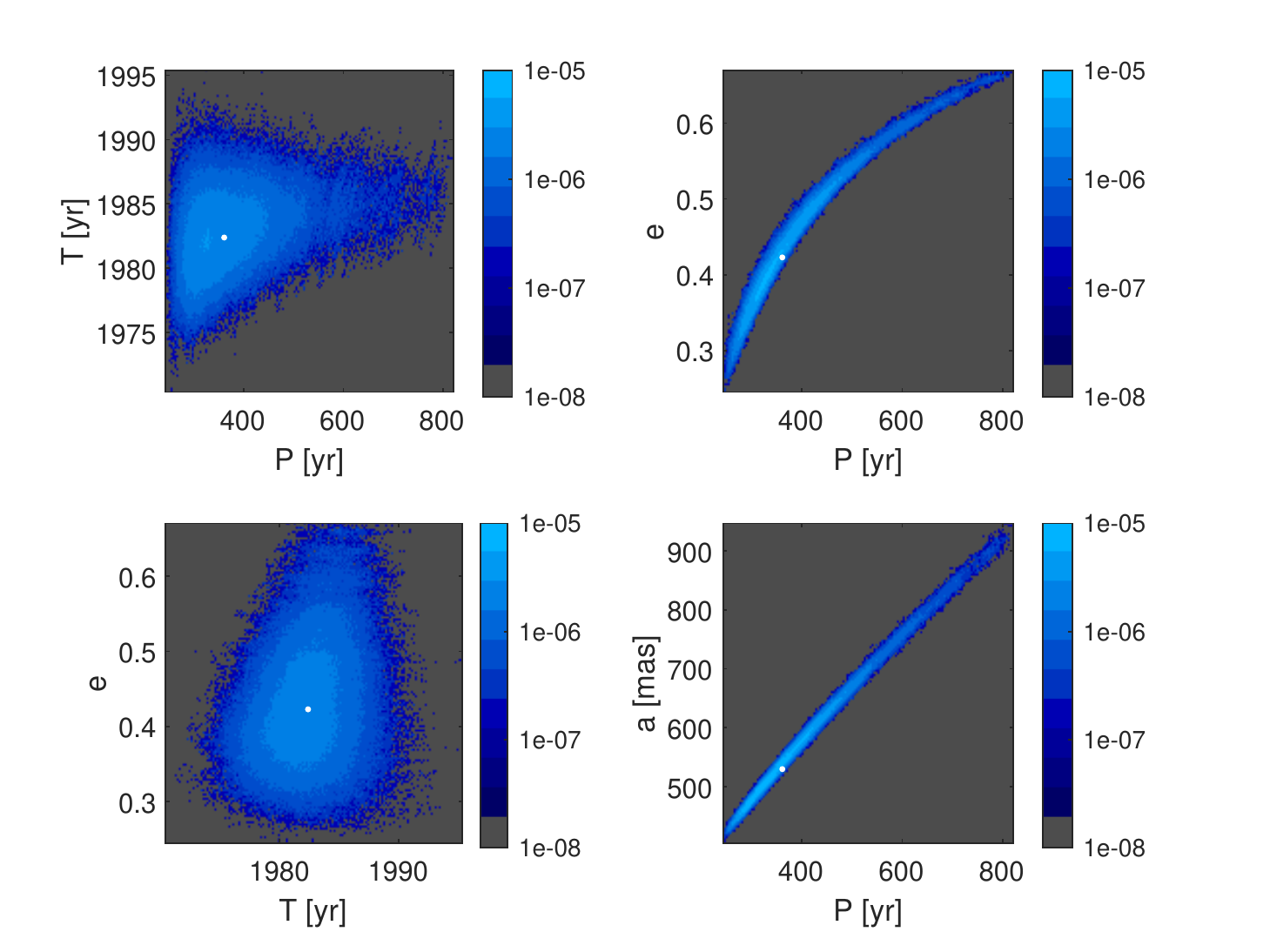}\\
\includegraphics[width=.99\textwidth]{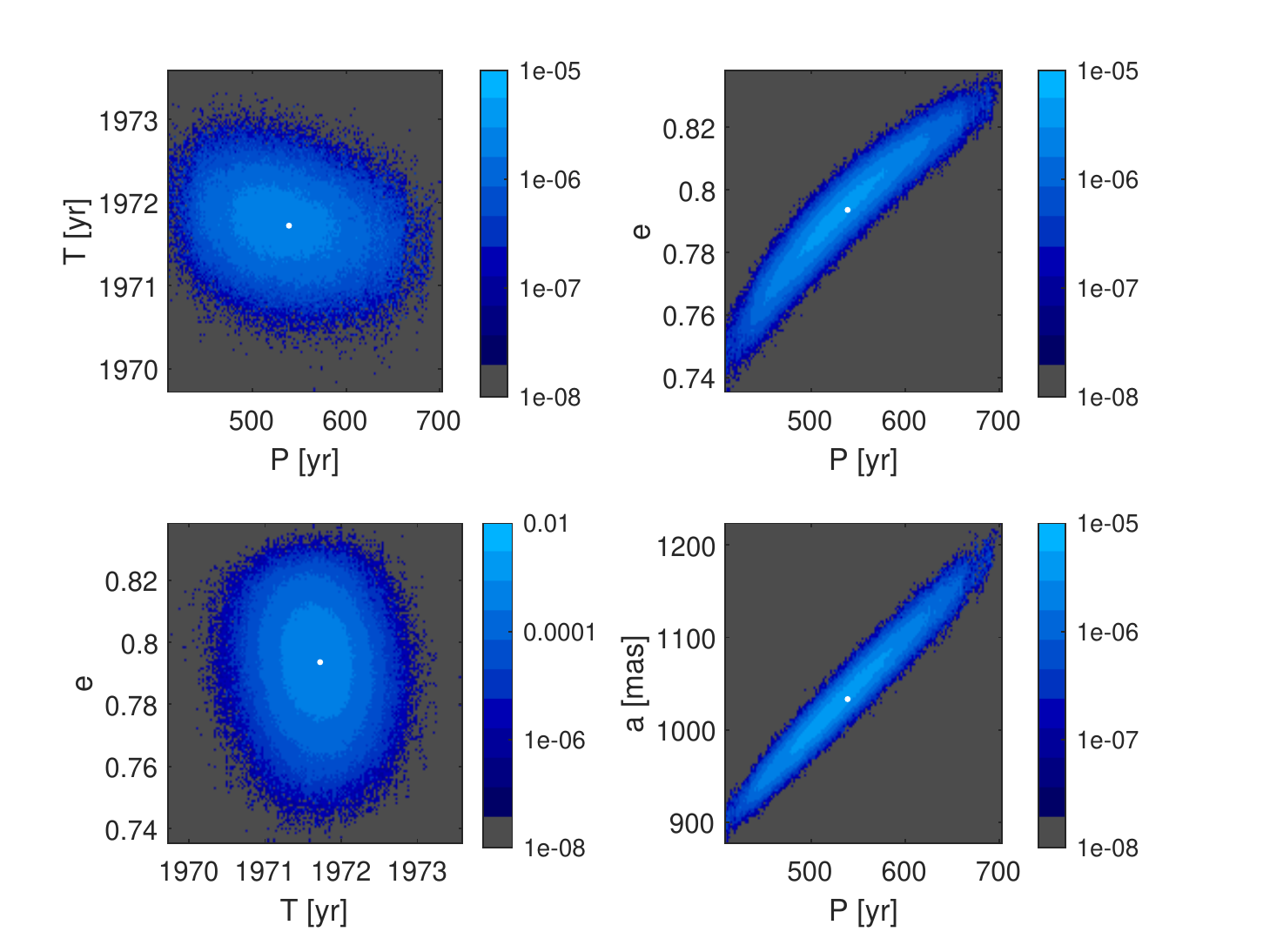}\\
\includegraphics[width=.99\textwidth]{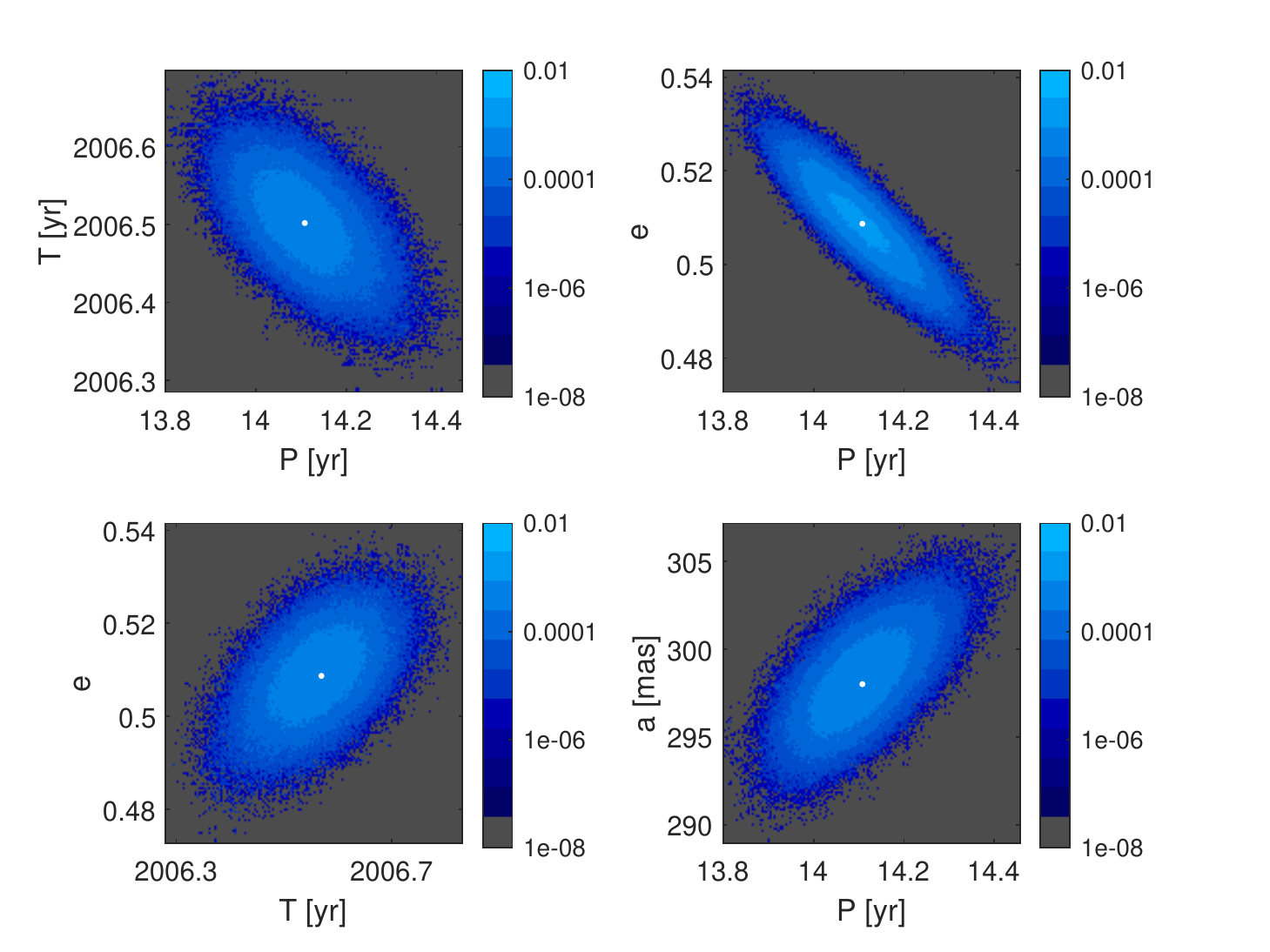}
\end{minipage} 
\caption{Maximum likelihood orbits and joint PDF estimates for three representative cases. From top to bottom: BU220 (rather uncertain orbit), I292 (better defined orbit), YSC62 (well defined orbit). The size and color of the dots indicate the weight (uncertainty) of each observation; large clearer dots indicate larger errors (smaller dots are from more recent interferometric measurements, including - but not limited - to our own). The green line indicates the line of nodes, while the black line indicates the direction to the periastron. In the PDF level curves (two columns on the right), the white dot indicates the MAP value given in Table~\ref{tab:orbel1}.} \label{fig:examples}
\end{figure}

In scenarios with highly disperse and asymmetric probability densities, the expected value rarely provides a meaningful estimate of the parameters, typically yielding orbits that are not in good agreement with the observations. Maximum likelihood (ML) and Maximum a posteriori (MAP) estimates--defined as the values that maximize the likelihood function and the posterior distribution, respectively \citep{Gelmanet2013} -- are preferred in those situations. In this work, the likelihood functions assumes a Gaussian model for observation, and the priors are just uniform densities in a certain interval (e.g., the prior of eccentricity $p(e)$ is 1 in [0,1) and 0 out of that range). In orbits with good orbital coverage, e.g., CVN16AaAb, FIN297AB, and FIN370, YSC62 (also Section~\ref{sec:sb2}), the expected value approximately coincides with the MAP/ML estimate, see Figure~\ref{fig:examples}.


\floattable
\rotate
\begin{deluxetable}{ccccccccccc}
\tablecaption{Orbital elements of our visual binaries.} \label{tab:orbel1}
\tabletypesize{\footnotesize}
\tablecolumns{11}
\tablewidth{0pt}
\tablehead{
\colhead{WDS} &
\colhead{Discoverer} &
\colhead{P} &
\colhead{T$_0$} &
\colhead{e} &
\colhead{a} &
\colhead{$\omega$} &
\colhead{$\Omega$} &
\colhead{i} &
\colhead{Gr} &
\colhead{Orbit}\\
\colhead{} &
\colhead{designation} &
\colhead{(yr)} &
\colhead{(yr)} &
\colhead{} &
\colhead{(mas)} &
\colhead{($^{\circ}$)} &
\colhead{($^{\circ}$)} &
\colhead{($^{\circ}$)} &
\colhead{Current\tablenotemark{a} $\rightarrow$ New} & \colhead{reference\tablenotemark{a}}
}
\startdata
10043$-$2823 & I292 & $539$ &$1971.72$ &$0.7936$ &$1033$ &$74.7$ &$143.0$ &$139.6$ & 5$\rightarrow$5 & USN2000b \\ 
& & $536_{-30}^{+33}$ &$1971.72_{-0.25}^{+0.24}$ &$0.7927_{-0.0096}^{+0.0093}$ &$1030_{-32}^{+34}$ &$74.6_{-1.2}^{+1.2}$ &$142.8_{-1.1}^{+1.1}$ &$139.5_{-1.1}^{+1.1}$ &  &\\ 
10174$-$5354 & CVN16AaAb & $5.320$ &$2005.948$ &$0.1350$ &$97.63$ &$107.1$ &$119.2$ &$19.9$ & 2$\rightarrow$2 & Tok2019c \\ 
& & $5.320_{-0.007}^{+0.007}$ &$2005.947_{-0.034}^{+0.035}$ &$0.1348_{-0.0063}^{+0.0064}$ &$97.58_{-0.25}^{+0.27}$ &$107.4_{-2.2}^{+2.6}$ &$118.9_{-2.4}^{+2.1}$ &$19.8_{-1.2}^{+1.2}$ &  &\\
11125$-$1830 & BU220 & $361$ &$1982.4$ &$0.423$ &$530$ &$155.5$ &$145.30$ &$99.77$ & 4$\rightarrow$4 & Tok2015c \\
& & $358_{-34}^{+50}$ &$1982.6_{-1.7}^{+1.7}$ &$0.420_{-0.041}^{+0.049}$ &$526_{-35}^{+49}$ &$155.9_{-3.3}^{+3.2}$ &$145.27_{-0.30}^{+0.30}$ &$99.78_{-0.22}^{+0.20}$ &  &\\
13145$-$2417 & FIN297AB & $60.61$ &$1957.50$ &$0.7323$ &$246.2$ &$111.22$ &$10.74$ &$67.58$ & 2$\rightarrow$2 & Tok2018i \\ 
& & $60.52_{-0.27}^{+0.25}$ &$1957.59_{-0.25}^{+0.27}$ &$0.7324_{-0.0049}^{+0.0049}$ &$245.9_{-1.3}^{+1.2}$ &$111.32_{-0.34}^{+0.35}$ &$10.75_{-0.33}^{+0.33}$ &$67.54_{-0.15}^{+0.14}$ &  &\\ 
13574$-$6229 & FIN370 & $18.768$ &$1968.35$ &$0.2185$ &$135.69$ &$181.4$ &$90.1$ &$144.40$ & 2$\rightarrow$2  & Msn2019 \\ 
& & $18.770_{-0.043}^{+0.042}$ &$1968.34_{-0.12}^{+0.11}$ &$0.2187_{-0.0054}^{+0.0055}$ &$135.74_{-0.89}^{+0.88}$ &$181.1_{-3.2}^{+3.2}$ &$90.0_{-1.5}^{+1.5}$ &$144.33_{-0.85}^{+0.88}$ &  &\\
15160$-$0454 & STF3091AB & $148.5$ &$1880.26$ &$0.754$ &$670$ &$284.8$ &$46.03$ &$92.06$ & 2$\rightarrow$2  & Msn2019 \\ 
& & $148.8_{-1.1}^{+1.2}$ &$1880.29_{-0.68}^{+0.68}$ &$0.740_{-0.035}^{+0.036}$ &$656_{-29}^{+38}$ &$285.2_{-1.3}^{+1.3}$ &$46.08_{-0.15}^{+0.14}$ &$92.10_{-0.11}^{+0.10}$ &  &\\ 
15420$+$0027 & A2176 & $53.16$ &$1934.04$ &$0.6305$ &$140.1$ &$258.9$ &$120.2$ &$19.7$ & 2$\rightarrow$2 & Tok2015c \\ 
& & $53.13_{-0.66}^{+0.69}$ &$1934.06_{-0.67}^{+0.63}$ &$0.6298_{-0.0073}^{+0.0074}$ &$140.1_{-1.8}^{+1.9}$ &$258.8_{-7.3}^{+8.0}$ &$120.7_{-9.2}^{+8.0}$ &$20.0_{-2.6}^{+2.3}$ & &\\ 
17005$+$0635 & CHR59 & $26.472$ &$1988.57$ &$0.0814$ &$191.29$ &$277.2$ &$67.94$ &$66.89$ & 4$\rightarrow$4 & Rbr2018 \\ 
& & $26.469_{-0.059}^{+0.059}$ &$1988.56_{-0.18}^{+0.17}$ &$0.0801_{-0.0070}^{+0.0070}$ &$191.28_{-0.38}^{+0.37}$ &$277.0_{-2.4}^{+2.3}$ &$67.99_{-0.28}^{+0.29}$ &$66.89_{-0.12}^{+0.12}$ &  &\\ 
17077$+$0722 & YSC62 & $14.108$ &$2006.502$ &$0.5087$ &$298.0$ &$200.72$ &$61.01$ &$114.46$ & 3$\rightarrow$2 & Msn2018a \\ 
& & $14.113_{-0.052}^{+0.053}$ &$2006.501_{-0.033}^{+0.032}$ &$0.5084_{-0.0051}^{+0.0051}$ &$298.2_{-1.4}^{+1.4}$ &$200.74_{-0.78}^{+0.79}$ &$61.01_{-0.20}^{+0.20}$ &$114.43_{-0.30}^{+0.30}$ &  &\\
17155$+$1052 & HDS2440 & $25.73$ &$2002.47$ &$0.252$ &$153.7$ &$109.3$ &$75.25$ &$120.5$ & 3$\rightarrow$2 & Cve2014 \\ 
& & $25.71_{-0.26}^{+0.26}$ &$2002.46_{-0.12}^{+0.12}$ &$0.253_{-0.012}^{+0.012}$ &$153.7_{-1.2}^{+1.3}$ &$109.1_{-1.5}^{+1.5}$ &$75.33_{-0.59}^{+0.60}$ &$120.5_{-1.0}^{+1.0}$ &  &\\ 
17181$-$3810 & SEE324 & $93$ &$2020.8$ &$0.396$ &$190.6$ &$359$ &$84.6$ &$118.1$ & 5$\rightarrow$5 & Tok2015c \\ 
& & $93_{-11}^{+15}$ &$2021.0_{-1.8}^{+1.3}$ &$0.413_{-0.035}^{+0.037}$ &$194.1_{-7.0}^{+8.9}$ &$305_{-289??}^{+38}$ &$83.9_{-2.2}^{+1.9}$ &$117.7_{-2.8}^{+2.9}$ &  &\\ 
17283$-$2058  & A2244AB & $44.79$ &$1925.81$ &$0.5487$ &$170.48$ &$196.60$ &$92.75$ &$42.09$ & 2$\rightarrow$2 & Tok2019h \\
& & $44.81_{-0.19}^{+0.20}$ &$1925.77_{-0.41}^{+0.40}$ &$0.5488_{-0.0030}^{+0.0030}$ &$170.45_{-0.57}^{+0.57}$ &$196.60_{-0.89}^{+0.88}$ &$92.73_{-0.63}^{+0.60}$ &$42.07_{-0.39}^{+0.38}$ &  &\\ 
17571$+$0004 & STF2244 & $507$ &$1977.63$ &$0.560$ &$1048$ &$288.4$ &$98.25$ &$83.58$ & 3$\rightarrow$3 & Msn2017e \\
& & $502_{-23}^{+34}$ &$1977.71_{-0.62}^{+0.67}$ &$0.555_{-0.022}^{+0.028}$ &$1044_{-23}^{+33}$ &$288.6_{-1.4}^{+1.4}$ &$98.27_{-0.14}^{+0.13}$ &$83.60_{-0.17}^{+0.15}$ &  &\\ 
19190$-$3317 & I253AB & $56.59$ &$1939.36$ &$0.738$ &$448$ &$141.7$ &$137.96$ &$93.31$ & 4$\rightarrow$4  & B\_\_1954 \\ 
& & $56.60_{-0.36}^{+0.37}$ &$1939.35_{-0.51}^{+0.50}$ &$0.740_{-0.013}^{+0.016}$ &$449_{-18}^{+21}$ &$141.5_{-4.3}^{+4.3}$ &$137.95_{-0.16}^{+0.17}$ &$93.30_{-0.22}^{+0.21}$ &  &\\ 
19471$-$1953 & BU146 & $115.2$ &$1998.28$ &$0.8144$ &$1106$ &$278.92$ &$34.32$ &$75.81$ & 4$\rightarrow$4 & Msn2019\\ 
& & $115.1_{-1.6}^{+1.6}$ &$1998.29_{-0.13}^{+0.12}$ &$0.8149_{-0.0057}^{+0.0058}$ &$1107_{-16}^{+18}$ &$278.87_{-0.30}^{+0.30}$ &$34.35_{-0.25}^{+0.25}$ &$75.85_{-0.31}^{+0.32}$ &  & \\ 
\enddata
\tablenotetext{a}{Grades and references taken from the latest version of the USNO Orbit catalogue.}
\end{deluxetable}

The orbital coverage, and the reliability of the fitted orbital parameters, ranges from what one could consider as an almost final orbits (see previous paragraph), to those with very poor coverage and rather uncertain orbits (e.g., BU220, I292, SEE324, and STF2244). In the penultimate column we give a suggestion of the new "Grade" of the orbit as defined in the WDS, based on qualitative appraisal of the orbital coverage and the formal errors of the orbital elements. The last column gives the reference for the latest orbit published for each object according to the USNO orbit catalogue, using the nomenclature adopted in the WDS.\newline\newline

In Figure~\ref{fig:examples} we show examples of orbital solutions and PDFs for objects in our sample. Looking at Table~\ref{tab:orbel1} and Figure~\ref{fig:examples} it can be seen that, in general, well determined orbits show a ML value that approximately coincides with the 2nd quartile of the PDF, the inter-quartile range is relatively well constrained, and the PDFs show a Gaussian-like distribution; whereas poor orbits show PDFs with very long tails (and therefore large inter-quartile ranges) on which the ML value usually differs by much from the 2nd quartile, and the PDFs are tangled. For completeness, in the site \url{http://www.das.uchile.cl/~rmendez/0001_Research/2_Visual_Binaries/Mendez_Claveria_Costa_AJ_2020/} we make available orbital plots and the relevant PDFs for all systems in our sample.

\subsection{Mass sums and dynamical parallaxes}\label{sec:massum}

In Table~\ref{tab:para} we present dynamical parallaxes and mass sums for our visual binaries. The first column gives the WDS designation, the second column the trigonometric parallax from Gaia DR2 -unless otherwise noted, with its uncertainty in the second line; in the third column we give our dynamical parallax; in the fourth and fifth columns we give our dynamical masses for the primary and secondary components, respectively; in the sixth column the total dynamic mass, and finally in the last column the astrometric mass sum calculated from the published parallax and the period and semi-major axis from Table~\ref{tab:orbel1}. With the exception of the second column, for each object in the upper line we indicate the mass sum from the maximum likelihood (ML) solution, while in the second line we report the second quartile (median), with the first quartile as a subscript and the third quartile as a superscript\footnote{These quartiles come from the MCMC orbital results alone, i.e., they do not include the error in the trigonometric parallax.}. A few of our objects do not have a Gaia parallax, their distance uncertainties being larger. There seems to be a slight tendency of these objects at having larger separations (mean of 0.5~arcsec) in comparison with the ones that do have a Gaia parallax, and also at having inclinations closer to 90~deg (i.e., nearly edge on). This is probably an indication of the observational difficulties faced by the Gaia satellite at resolving systems close to its angular resolution limit. It is well known that Hipparcos parallaxes were indeed biased due to the orbital motion of the binary (i.e., the parallax and orbit signal are blended), as shown by \cite{Soder1999} (see, in particular his Section~3.1, and Table~2), and it is likely that Gaia will suffer from a similar problem\footnote{For example, according to Tokovinin's multiple star catalogue, HIP 64421 contains a binary with a 27~yr orbit. Its Hipparcos parallax is 8.6~mas, its dynamical parallax is 8.44~mas, and its Gaia DR2 parallax is 3~mas. However, Gaia does give a consistent parallax for the C component (at 1.9~arcsec): $9.7 \pm 0.3$~mas, see \url{http://www.ctio.noao.edu/~atokovin/stars/}. There are other examples like this in the cited catalog.}. During each observation Gaia is not expected to resolve systems closer than about 0.4~arcsec, though over the mission there will be a final resolution of 0.1~arcsec. This is shown graphically in Figure~1 from \citet{Ziegleret2018}, where the current resolution of the second Gaia data release is shown to be around 1~arcsec (it is a function of the magnitude difference between primary and secondary\footnote{It is expected that, from the third Gaia data release and on, the treatment of binary stars will be much improved, by incorporating orbital motion (and its impact on the photocenter position of unresolved pairs) into the overall astrometric solution, thus suppressing/alleviating the parallax bias significantly, this in turns calls precisely for having good orbital elements for these binaries, which is one of the secondary goals of our project.}).

\floattable
\begin{deluxetable}{cccccccc}
\tablecaption{Trigonometric and dynamical parallaxes (visual binaries)\tablenotemark{a}.} 
\label{tab:para}
\tablecolumns{8}
\tablewidth{0pt}
\tablehead{
\colhead{WDS} &
\colhead{Discoverer} &
\colhead{Trig. Parallax} &
\colhead{Dyn. parallax} & \colhead{Mass$^{\mbox{\tiny{dyn}}}_{\mbox{\tiny{P}}}$} & \colhead{Mass$^{\mbox{\tiny{dyn}}}_{\mbox{\tiny{S}}}$} & \colhead{Mass$^{\mbox{\tiny{dyn}}}_{\mbox{\tiny{T}}}$} & \colhead{Mass$_{\mbox{\tiny{T}}}$\tablenotemark{b}}\\
\colhead{} &
\colhead{designation} &
\colhead{(mas)} &
\colhead{(mas)} & 
\colhead{($\it{M}_{\odot}$)} & 
\colhead{($\it{M}_{\odot}$)} & 
\colhead{($\it{M}_{\odot}$)} & 
\colhead{($\it{M}_{\odot}$)}
}
\startdata
10043$-$2823 & I292 & $10.07$ &$11.11$ &$1.470$ &$1.300$ &$2.771$ &$3.73$\\ 
& & $\pm0.34$ &$11.12_{-0.13}^{+0.13}$ &$1.469_{-0.007}^{+0.007}$ &$1.300_{-0.006}^{+0.006}$ &$2.769_{-0.013}^{+0.013}$ &$3.73_{-0.11}^{+0.11}$\\ 
10174$-$5354 & CVN16AaAb & $51.00$ &$45.13$ &$0.188$ &$0.170$ &$0.358$ &$0.248$\\ 
& & $\pm0.30$ &$45.11_{-0.16}^{+0.17}$ &$0.188_{-0.001}^{+0.001}$ &$0.170_{-0.000}^{+0.000}$ &$0.358_{-0.001}^{+0.001}$ &$0.247_{-0.002}^{+0.002}$\\ 
11125$-$1830 & BU220 & $6.50$ &$6.474$ &$2.558$ &$1.651$ &$4.208$ &$4.159$\\ 
& & $\pm0.71$\tablenotemark{c} &$6.476_{-0.052}^{+0.053}$ &$2.557_{-0.010}^{+0.010}$ &$1.651_{-0.006}^{+0.006}$ &$4.208_{-0.016}^{+0.016}$ &$4.162_{-0.085}^{+0.087}$\\ 
13145$-$2417 & FIN297AB & $11.22$ &$10.919$ &$1.565$ &$1.556$ &$3.121$ &$2.874$\\ 
& & $\pm0.54$ &$10.914_{-0.041}^{+0.041}$ &$1.566_{-0.002}^{+0.002}$ &$1.556_{-0.002}^{+0.002}$ &$3.122_{-0.005}^{+0.005}$ &$2.871_{-0.028}^{+0.028}$\\ 
13574$-$6229 & FIN370 & $8.06$ &$13.61$ &$1.459$ &$1.356$ &$2.815$ &$13.55$\\ 
& & $\pm0.92$ &$13.61_{-0.11}^{+0.10}$ &$1.459_{-0.005}^{+0.005}$ &$1.356_{-0.004}^{+0.004}$ &$2.815_{-0.009}^{+0.009}$ &$13.57_{-0.27}^{+0.27}$\\ 
15160$-$0454 &STF3091AB & $16.0$ &$18.13$ &$1.216$ &$1.069$ &$2.285$ &$3.33$\\ 
& & $\pm1.5$\tablenotemark{c} &$17.67_{-0.97}^{+1.26}$ &$1.228_{-0.032}^{+0.027}$ &$1.079_{-0.027}^{+0.023}$ &$2.307_{-0.059}^{+0.050}$ &$3.11_{-0.43}^{+0.62}$\\ 
15420+0027 & A2176 & $6.12$ &$6.510$ &$1.762$ &$1.762$ &$3.523$ &$4.24$\\ 
& & $\pm0.74$\tablenotemark{c} &$6.519_{-0.062}^{+0.063}$ &$1.761_{-0.007}^{+0.007}$ &$1.761_{-0.007}^{+0.007}$ &$3.523_{-0.015}^{+0.015}$ &$4.26_{-0.10}^{+0.11}$\\ 
17005+0635 & CHR59 & $14.10$ &$15.929$ &$1.584$ &$0.888$ &$2.471$ &$3.565$\\ 
& & $\pm0.16$ &$15.930_{-0.029}^{+0.028}$ &$1.584_{-0.001}^{+0.001}$ &$0.888_{-0.001}^{+0.001}$ &$2.471_{-0.002}^{+0.002}$ &$3.565_{-0.017}^{+0.017}$\\ 
17077+0722 & YSC62 & $82.00$ &$78.81$ &$0.145$ &$0.127$ &$0.272$ &$0.241$\\ 
& & $\pm2.30$\tablenotemark{c} &$78.83_{-0.34}^{+0.34}$ &$0.145_{-0.000}^{+0.000}$ &$0.126_{-0.000}^{+0.000}$ &$0.272_{-0.000}^{+0.000}$ &$0.241_{-0.003}^{+0.003}$\\ 
17155+1052 &HDS2440 & $13.50$ &$14.15$ &$1.030$ &$0.904$ &$1.934$ &$2.227$\\ 
& & $\pm0.15$ &$14.17_{-0.13}^{+0.14}$ &$1.030_{-0.004}^{+0.003}$ &$0.904_{-0.003}^{+0.003}$ &$1.934_{-0.007}^{+0.006}$ &$2.232_{-0.054}^{+0.058}$\\ 
17181$-$3810 & SEE324 & $5.11$ &$5.90$ &$2.034$ &$1.874$ &$3.91$ &$6.0$\\ 
& & $\pm0.17$ &$5.90_{-0.44}^{+0.53}$ &$2.034_{-0.076}^{+0.073}$ &$1.873_{-0.069}^{+0.066}$ &$3.91_{-0.15}^{+0.14}$ &$6.0_{-1.1}^{+1.5}$\\ 
17283$-$2058 & A2244AB & $10.68$ &$9.894$ &$1.301$ &$1.250$ &$2.551$ &$2.027$\\ 
& & $\pm0.11$ &$9.888_{-0.068}^{+0.068}$ &$1.301_{-0.003}^{+0.004}$ &$1.250_{-0.003}^{+0.003}$ &$2.551_{-0.007}^{+0.007}$ &$2.024_{-0.036}^{+0.036}$\\ 
17571+0004 & STF2244 & $8.44$ &$10.69$ &$1.896$ &$1.774$ &$3.670$ &$7.47$\\ 
& & $\pm0.73$\tablenotemark{c} &$10.72_{-0.15}^{+0.13}$ &$1.895_{-0.010}^{+0.012}$ &$1.772_{-0.010}^{+0.011}$ &$3.667_{-0.020}^{+0.023}$ &$7.50_{-0.27}^{+0.24}$\\ 
19190$-$3317 & I253AB & $18.2$ &$23.6$ &$1.197$ &$0.927$ &$2.124$ &$4.69$\\ 
& & $\pm1.1$\tablenotemark{c} &$23.7_{-1.1}^{+1.3}$ &$1.196_{-0.025}^{+0.023}$ &$0.926_{-0.017}^{+0.016}$ &$2.122_{-0.042}^{+0.039}$ &$4.72_{-0.60}^{+0.74}$\\ 
19471$-$1953 & BU146 & $5.44$ &$40.94$ &$0.791$ &$0.694$ &$1.485$ &$634$\\ 
& & $\pm0.99$ &$40.99_{-0.47}^{+0.51}$ &$0.791_{-0.003}^{+0.003}$ &$0.694_{-0.003}^{+0.002}$ &$1.485_{-0.006}^{+0.005}$ &$636_{-19}^{+21}$\\ 
\enddata
\tablenotetext{a}{Values smaller than $5\times10^{-4}$ are reported as 0.000.}
\tablenotetext{b}{Using the solution from Table~\ref{tab:orbel1}, and the published trigonometric parallax indicated on the second column of this table.}
\tablenotetext{c}{No Gaia parallax. Value from Hipparcos.}
\end{deluxetable}

To calculate the dynamical parallaxes given in this table as a consistency check, we have adopted photometric values for the primary ($V_{\mbox{P}}$) and secondary ($V_{\mbox{S}}$) from Table~\ref{tab:photom1}, the values of $P$ and $a$ from Table~\ref{tab:orbel1}, and the MLR for main sequence stars from \cite{HenMcC1993}, who provide polynomial relationships between mass {\it vs.} $M_V$ for objects less massive than about 1$M_{\odot}$\footnote{Several of our objects on Table~\ref{tab:para} have masses above 1$M_{\odot}$, but the polynomial fits of the MLR are gentle enough to allow some extrapolation, see, e.g., Figure~2 on \cite{HenMcC1993}.}. The quoted uncertainty values for the dynamical parallax come exclusively from the range of solutions of our MCMC simulations, and not from uncertainties on either the photometry (which can introduce a major uncertainty on the derived dynamical parallax and the implied dynamical masses), nor the observational (intrinsic or not) width of the MLR.

\begin{figure}
\plotone{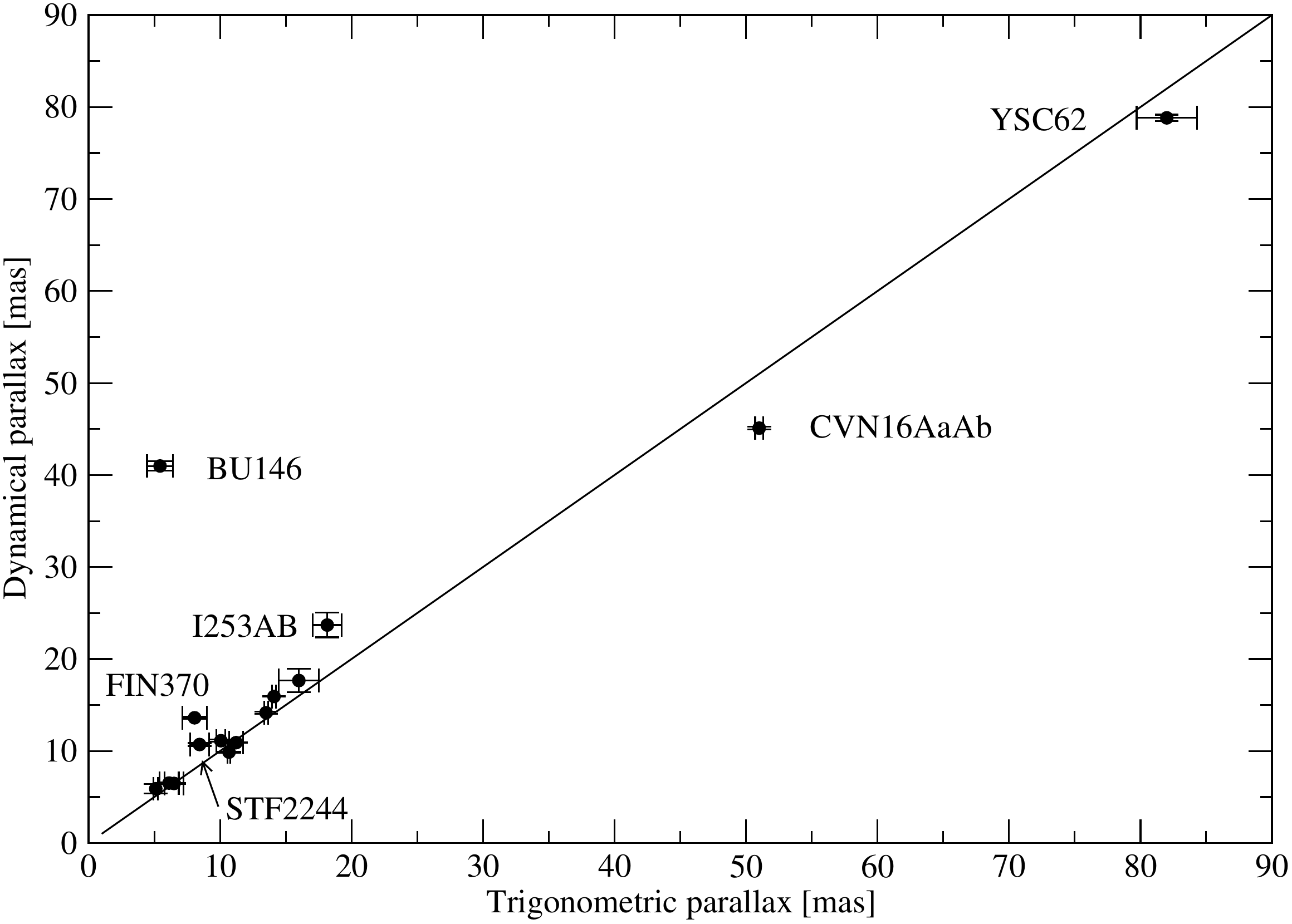}
\caption{Comparison of the dynamical and trigonometric parallaxes of the objects included in Table~\ref{tab:para}. Using the Discoverer designation, we have highlighted those objects that exhibit a very large difference between their dynamical and astrometric parallaxes, or that appear as discrepant in Figure~\ref{fig:mass}. The solid line is not a fit; it only shows the expected one-to-one relationship. In the ordinate, the quantity shown is the $2^{nd}$ quartile from Table~\ref{tab:para}.\label{fig:para}}
\end{figure}

\begin{figure}
\plotone{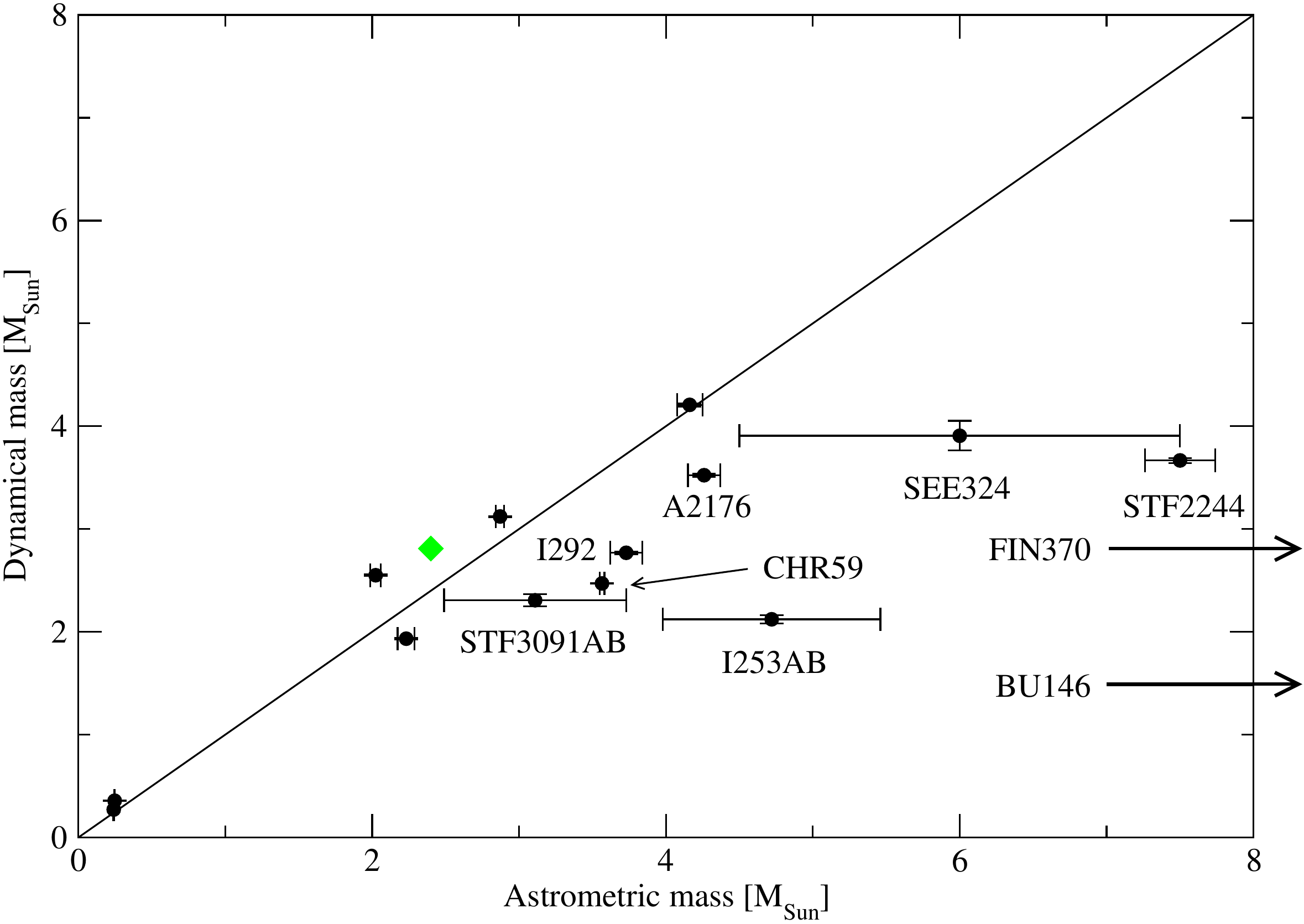}
\caption{Comparison of the dynamical and astrometric mass sums of the objects included in  Table~\ref{tab:para}. Using the Discoverer designation, we have highlighted those objects that exhibit a very large difference between their dynamical and astrometric masses. The solid line is not a fit; it only shows the expected one-to-one relationship. In both axes the quantity shown is the $2^{nd}$ quartile from Table~\ref{tab:para}. BU146's astrometric mass falls out of scale to the right, so we depict it with an arrow at the level of its dynamical mass (see Section~\ref{sec:hr} for further details). This is the same situation with FIN370 adopting the Gaia parallax, however, if we adopt the Hipparcos parallax our orbit will lead to a much smaller mass, indicated by the green diamond (for further details see Section~\ref{sec:hr}). \label{fig:mass}}
\end{figure}

In Figures~\ref{fig:para} and \ref{fig:mass} we show the values of Table~\ref{tab:para} in graphical form. Generally speaking, there is good agreement between the dynamical and astrometric parallaxes and masses. There are some notable exceptions that can be attributed to either a poor orbit determination, a large parallax uncertainty, poor photometry, or a combination of these. This is discussed on an object-by-object basis in more detail in Section~\ref{sec:hr}. Also, note that a few objects in our list (A2176, FIN370, and SEE324, individualized on the figures) are not on the main sequence. For them the adopted MLR -and therefore, the implied dynamical parallax- is not valid (the extreme case being FIN370, further discussed in Section~\ref{sec:hr}), see also Figure~\ref{fig:hrdiag}. 

Considering that dynamical parallaxes are derived from photometry, they are prone to being systematically overestimated due to interstellar absorption. Although our sample encompasses relatively small heliocentric distance (our most distant target is SEE324 at almost 196~pc, followed by BU146 at 184~pc), this particular dataset of binaries is however located towards the inner part of the Milky Way, and at relatively low Galactic latitudes, where extinction is generally larger than at high Galactic latitude. We have used the reddening model by \citet{MenVan1998} to estimate the extinction for all our targets and evaluate its impact on our dynamical parallaxes. For the most distant binary, SEE324, the predicted extinction in the $V$-band is 0.18~mag at that Galactic location. With this extinction, the ML dynamical parallax changes from 5.90 (no extinction) to 5.82~mas (extincted), i.e., completely within the computed inter-quartile range reported in Table~\ref{tab:para}. However, given the patchy distribution of dust along the line-of-sight, our most extincted target according to the \citet{MenVan1998} reddening model is not SEE324, but instead it is FIN370 at a distance of 124~pc, located towards $(l,b)=(310\degr.5, -0\degr.6)$, with $A_V = 0.38$~mag. In this case, the corresponding ML dynamical parallax changes from 13.61~mas (no extinction) to 13.23~mas (extincted), a difference of only 0.38~mas, which is more than two times smaller than the quoted uncertainty for the trigonometric parallax of this target. After FIN370, and slightly more extincted than SEE324, we have STF2244, with $A_V = 0.21$~mag. In this case, the difference between the dynamical parallax with and without extinction correction is within $1 \sigma$ of the inter-quartile range for this object given in Table~\ref{tab:para} (10.69~mas with no extinction to 10.52~mas extincted). For all other targets in our list, the effect of interstellar absorption effects in the calculation of the dynamical parallaxes is minimal, and do not change in any significant way the values given in Table~\ref{tab:para}.

\subsection{The double line spectroscopic binary YSC8 (WDS15006+0836)}\label{sec:sb2}

One of our Speckle targets, YSC8, is a double-line spectroscopic binary (SB2 hereafter). For this object, we performed a joint solution of the astrometric and radial velocity data using our MCMC code as described in Section~3.6 of \cite{Mendezetal2017}. We retrieved its radial velocity curve from the 9th Catalogue of Spectroscopic Binary Orbits (\cite{Pouret2004})\footnote{Updated regularly, and available at \url{https://sb9.astro.ulb.ac.be/}}, and we combined it with our HRCam astrometric measurements, and with historical data from the WDS catalogue. We note that ours is the first combined orbit computed for this system.

The resulting orbital elements are presented in the first seven columns of Table~\ref{tab:orbelYSC8}. That table also includes the systemic radial velocity (column 8), the mass ratio with the derived uncertainties (column 9), and the dynamically self-consistent orbital parallax (column 10).

\begin{figure}
\plotone{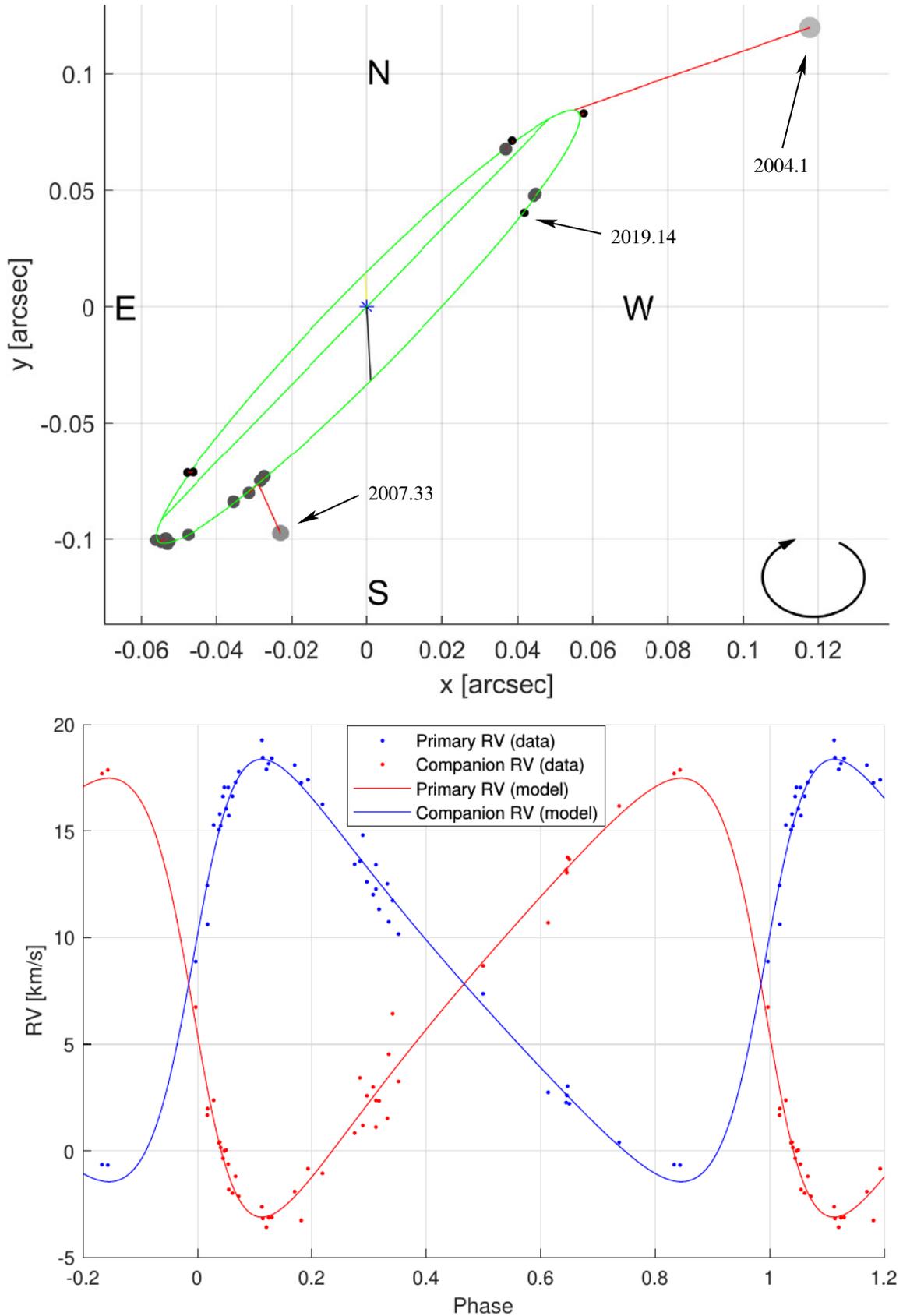}
\caption{Results of our MCMC simultaneous fits to YSC8. The upper panel shows the data points, and the astrometric orbit, together with the line of nodes and the direction to the periastron. The lower panel shows the radial velocity curve for both components. In the astrometric orbit, the two most-deviant points -which have been given a lower weight in the solution- are from CCD Speckle observations at WIYN by Horch and collaborators; specifically their first epoch on 2004.10 \citet{Horchet2008}, and their third epoch on 2007.33 \citet{Horchet2008}. We further note that all our SOAR/HRCam points exhibit very small residuals. Our last HRCam epoch at 2019.14 is indicated. The radial velocity amplitudes clearly point to an almost equal-mass system.\label{fig:ysc8}}
\end{figure}

\begin{figure}
\plotone{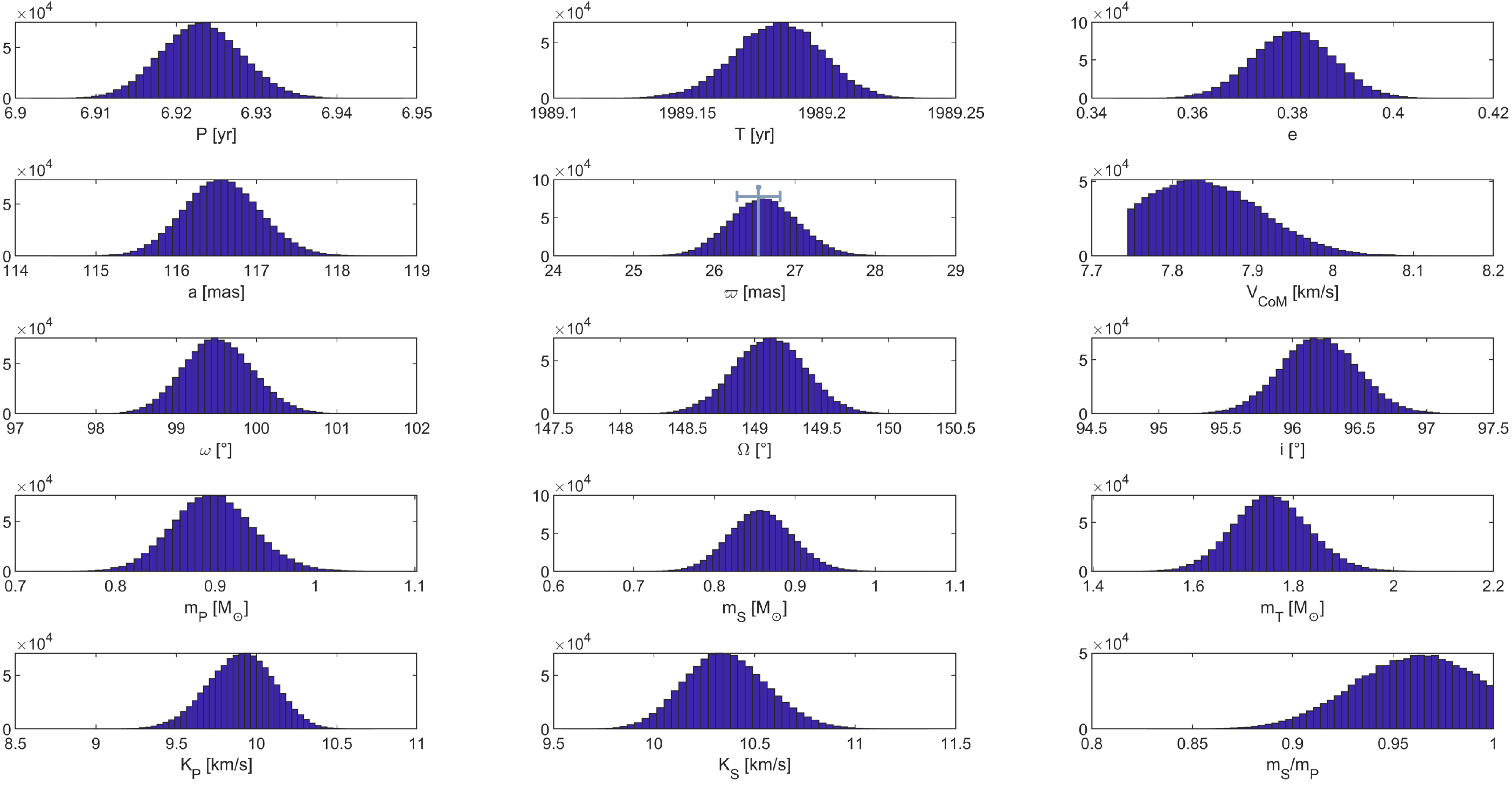}
\caption{Posterior distributions of the classical seven orbital elements, plus the (fitted) orbital parallax, together with the Gaia DR2 trigonometric parallax and its $\pm 1 \sigma$ error, the systemic velocity, the velocity amplitudes for both components, the mass sum, the mass ratio, and the individual component masses. The cut on the lower end of the $V_{\mbox{\tiny{CoM}}}$ histogram is a consequence of the fact that we discard solutions with $m_{\mbox{\tiny{S}}}/m_{\mbox{\tiny{P}}} > 1$. However, the method still provides a good estimate of the orbit and the distribution of orbital parameters. \label{fig:histYSC8}}
\end{figure}

Because the posterior PDFs obtained here are tighter and more Gaussian-like than those obtained for our visual binaries, the expected value offers a good estimate of the target parameter vector and thus will be the estimator of choice in this section (see Section~\ref{sec:orbel}). In Figure~\ref{fig:ysc8} we show the joint fit to the astrometric orbit (upper panel) and the radial velocity curve (lower panel). As it can bee seen from the table and figure, even though the astrometric orbit does not have an excellent phase coverage, the combined solution produces very precise orbital parameters. This is also evident in Figure~\ref{fig:histYSC8}, where we present the posterior PDFs, which exhibit tight and well-constrained distributions. In particular, judging from the quartile ranges, we can see that the mass ratio is determined with a 2\% uncertainty, while the uncertainty on the mass sum is 3\% (see Table~\ref{tab:dynplxYSC8}). The formal uncertainty on the individual component masses is $\sim 0.03 M_\odot$.

The orbital elements given in SB9 for YSC8, computed from the spectroscopic orbit alone, are $P = 6.914 \pm 0.021$~yr, $e=0.387 \pm 0.015$ and $\omega=283.4 \pm 2.6$~deg\footnote{With an ambiguity of 180 deg.}, in good agreement with our results. A purely astrometric orbit was reported by Tokovinin at the IAU Commission G1 on Binary and multiple stars systems\footnote{Inf. Circ. No. 195, 1, June 2018, available at \url{https://www.usc.gal/astro/circulares/cir195.pdf}}. He finds $P = 6.94$~yr\footnote{The period was fixed, not fitted: Andrei Tokovinin, personal communication.}, $T=2016.879$, $e= 0.374$, $\Omega= 149.2$~deg, $a= 117$~mas, $i= 96.3$~deg, $\omega= 99.3$~deg, using as last astrometric epoch 2018.164 from HRCam (this is the orbit included in the USNO orbit catalogue\footnote{Curiously, in SIMBAD's links to external archives, this object is not listed as belonging to the WDS catalogue as is usually done for all visual binaries.}).
Our latest HRCam point is however from 2019.14 (see Figure~\ref{fig:ysc8}), and was published in \cite{Tokoet2020}. A comparison with our values given in Table~\ref{tab:orbelYSC8} shows a fair agreement with Tokovinin's purely astrometric orbit. Further analysis is prevented by the lack of errors for the orbital parameters reported in the cited Circular.

\floattable
\begin{deluxetable}{cccccccccc}
\tablecaption{Orbital elements for the double line binary YSC8. \label{tab:orbelYSC8}}
\tabletypesize{\footnotesize}
\tablecolumns{10}
\tablewidth{0pt}
\tablehead{
\colhead{P} & \colhead{T$_0$} &
\colhead{e} & \colhead{a} & \colhead{$\omega$} & \colhead{$\Omega$} &
\colhead{i} & \colhead{$V_{\text{CoM}}$} & \colhead{$m_2/m_1$} & \colhead{$\varpi$}\\
\colhead{(yr)} & \colhead{(yr)} &
 & \colhead{(mas)} & \colhead{($^{\circ}$)} & \colhead{($^{\circ}$)} & \colhead{($^{\circ}$)} & \colhead{(km/s)}&\colhead{$~$} & \colhead{(mas)}
}
\startdata
$6.924$ &$1989.184$ &$0.3795$ &$116.54$ &$99.58$ &$149.10$ &$96.22$ &$7.833$ &$0.962$ &$26.65$\\ 
$6.923_{-0.003}^{+0.003}$ &$1989.184_{-0.011}^{+0.011}$ &$0.3801_{-0.0054}^{+0.0054}$ &$116.56_{-0.32}^{+0.32}$ &$99.51_{-0.28}^{+0.29}$ &$149.11_{-0.18}^{+0.18}$ &$96.20_{-0.19}^{+0.19}$ &$7.843_{-0.043}^{+0.049}$ &$0.958_{-0.020}^{+0.018}$ &$26.61_{-0.28}^{+0.29}$\\
\enddata
\end{deluxetable}

\floattable
\begin{deluxetable}{ccccccccc}
\tablecaption{Trigonometric and dynamical parallaxes and mass estimates for YSC8.}\label{tab:dynplxYSC8}
\tabletypesize{\footnotesize}
\tablecolumns{9}
\tablewidth{0pt}
\tablehead{
\colhead{Trig. Parallax} & \colhead{Dyn. parallax} &
\colhead{Mass$^{\mbox{\tiny{dyn}}}_{\mbox{\tiny{P}}}$} & \colhead{Mass$^{\mbox{\tiny{dyn}}}_{\mbox{\tiny{S}}}$} & \colhead{Mass$^{\mbox{\tiny{dyn}}}_{\mbox{\tiny{T}}}$} & \colhead{Mass$^{\mbox{\tiny{comb}}}_{\mbox{\tiny{P}}}$} & \colhead{Mass$^{\mbox{\tiny{comb}}}_{\mbox{\tiny{S}}}$} & \colhead{Mass$^{\mbox{\tiny{comb}}}_{\mbox{\tiny{T}}}$} & \colhead{Mass$_{\mbox{\tiny{T}}}$}\\
\colhead{(mas)} & \colhead{(mas)} &
 & \colhead{$M_{\odot}$} & \colhead{$M_{\odot}$} & \colhead{$M_{\odot}$} & \colhead{$M_{\odot}$} & \colhead{$M_{\odot}$}&  \colhead{$M_{\odot}$} 
}
\startdata
26.55 & 25.787 &$1.109$ &$0.816$ &$1.926$ &$0.889$ &$0.856$ &$1.745$ &$1.765$\\ 
$\pm 0.27$ &$25.794_{-0.084}^{+0.085}$ &$1.109_{-0.001}^{+0.001}$ &$0.816_{-0.001}^{+0.001}$ &$1.925_{-0.002}^{+0.002}$ &$0.897_{-0.027}^{+0.027}$ &$0.857_{-0.026}^{+0.026}$ &$1.754_{-0.050}^{+0.051}$ &$1.766_{-0.015}^{+0.016}$\\
\enddata
\end{deluxetable}

In Table~\ref{tab:dynplxYSC8} we present a comparison of the masses for the system, as well as the individual component masses obtained from the joint fit of the orbit to the astrometric and radial velocity data shown in Table~\ref{tab:orbelYSC8}. The format of the table is similar to that of Table~\ref{tab:para} in that it includes, for comparison purposes, dynamical parallaxes (second column) and individual and total dynamical masses (third to fifth columns) calculated in the same way as described in Section~\ref{sec:massum} for the visual binaries. The ninth column gives the total mass using the orbital elements shown in Table~\ref{tab:orbelYSC8}, but adopting the published trigonometric parallax given in the first column of this table (and, assuming it has no error, see below). The sixth to eight columns give the individual masses and the total mass computed allowing the parallax of the system be a free parameter of the MCMC code (and whose MAP and quartile values are given in the tenth column of Table~\ref{tab:orbelYSC8}). We note that the quartiles on Mass$_{\mbox{\tiny{T}}}$ in Table~\ref{tab:dynplxYSC8} {\it do not}
include any contribution from trigonometric parallax errors (just as in Table~\ref{tab:para}), while
Mass$^{\mbox{\tiny{comb}}}_{\mbox{\tiny{P}}}$,
Mass$^{\mbox{\tiny{comb}}}_{\mbox{\tiny{S}}}$, and
Mass$^{\mbox{\tiny{comb}}}_{\mbox{\tiny{T}}}$,
being derived from MCMC simulations that have the parallax as a free parameter, {\it do} include the extra variance from this parameter. As expected, when we take into account the variance of the parallax in the overall fit, the inter-quartile range for the derived masses increases (in this case, by a factor of three). We make the point that comparing these quantities in general is not strictly fair, since Mass$_{\mbox{\tiny{T}}}$ includes uncertainties in $a$ and $P$ exclusively (while the parallax is incorporated from other sources, assuming it has no error), instead Mass$^{\mbox{\tiny{comb}}}_{\mbox{\tiny{T}}}$ includes, in addition, the uncertainty on the parallax, determined self-consistently from the solution.

The agreement between all estimates of the mass is good. The trigonometric Gaia DR2 parallax and our orbital parallax ($\sim26.6$~mas) agree quite well and have the same uncertainty, but both are larger by about 0.8~mas than the dynamical parallax (which is about $3\sigma$ below the parallax uncertainty), thus leading to a slightly larger total dynamical mass than that obtained by adopting either the trigonometric or orbital parallax, as it can be readily seen in Table~\ref{tab:dynplxYSC8}.

As it was mentioned in Section~\ref{sec:photom}, the WDS combined magnitudes agree quite well with the SIMBAD and Hipparcos magnitudes. However, the opposite is not true, in the sense that this check {\it does not} necessarily proves that the individual component magnitudes listed by WDS are correct: indeed, we note that the photometry on WDS for the primary and secondary indicate a magnitude difference $\Delta V = 1.9$ which is in strong discord with our SOAR measurements of $\Delta y = 0.13$ and $\Delta I = 0.27$ (see Table~\ref{tab:photom1}). Such a large value of $\Delta V$ reported by WDS is also inconsistent with the mass ratio close to 1.0. Furthermore, SB9 reports a luminosity ratio of 0.9 for this system, corresponding to a magnitude difference close to 0.11 mag. No $V$mag is reported for this objects in SIMBAD, but the Hipparcos value (7.26) compares well with $V_{\mbox{ASAS}}=7.237 \pm 0.027$, $V_{\mbox{APASS}}=7.50 \pm 0.11$, and $V_{\mbox{PD2010}}=7.29$ magnitudes shown in Table~\ref{tab:photom2}, which also agree with $V_{\mbox{Sys}}=7.33$ from WDS given in Table~\ref{tab:photom1}. If we adopt the WDS system magnitude, but a  $\Delta V = \Delta y = 0.13$, then $V_{\mbox{P}}=8.02$, and $V_{\mbox{S}}=8.15$. For these new values, the dynamical masses turn out to be Mass$^{\mbox{\tiny{dyn}}}_{\mbox{\tiny{P}}}=1.02$, Mass$^{\mbox{\tiny{dyn}}}_{\mbox{\tiny{S}}}=1.00$, and Mass$^{\mbox{\tiny{dyn}}}_{\mbox{\tiny{T}}}=2.02$, which is a bit larger than the values using the WDS individual component values, and discrepant with the orbital mass by almost $0.26~M_\odot$. This is puzzling, considering that the astrometric total mass is determined with a  precision more than five times better than such a discrepancy. Nevertheless, we believe that these "corrected" individual magnitudes are actually better than the ones published in the WDS, see Figure~\ref{fig:hrdiag}.

We finally note the good agreement for the triad ($V_{\mbox{\tiny{CoM}}}$, $K_{\mbox{\tiny{P}}}$, $K_{\mbox{\tiny{S}}}$) reported by the 9th Catalogue of Spectroscopic Binary Orbits for YSC8: ($7.82 \pm 0.11$, $9.92 \pm 0.19$, $10.26 \pm 0.22$)  km~s$^{-1}$, and our calculations: ($7.84 \pm 0.05$, $9.91 \pm 0.14$, $10.35 \pm 0.14$) km~s$^{-1}$ (see Figure~\ref{fig:histYSC8} for the derived PDFs of these three kinematic quantities). This is particularly interesting, since it validates our extension of the proposal by \cite{WriHow2009} to the case of binary stars, following the mathematical formalism developed in \cite{Mendezetal2017}. We emphasize that both, $K_{\mbox{\tiny{P}}}$ and $K_{\mbox{\tiny{S}}}$ are not fitted directly from our data, instead they are derived from the other orbital elements in a dynamically self-consistent way (see, e.g., Equations~(10) and~(11) in Appendix~A.1 on \citet{Mendezetal2017}\footnote{Precisely, this constrain allows us to compute an orbital parallax, since the radial velocity curve is distance independent, while the semi-major axis in physical units is related to the astrometric semi-major axis through the parallax.}).


\section{HR diagram and comments on individual objects}\label{sec:hr}

\begin{figure}
\plotone{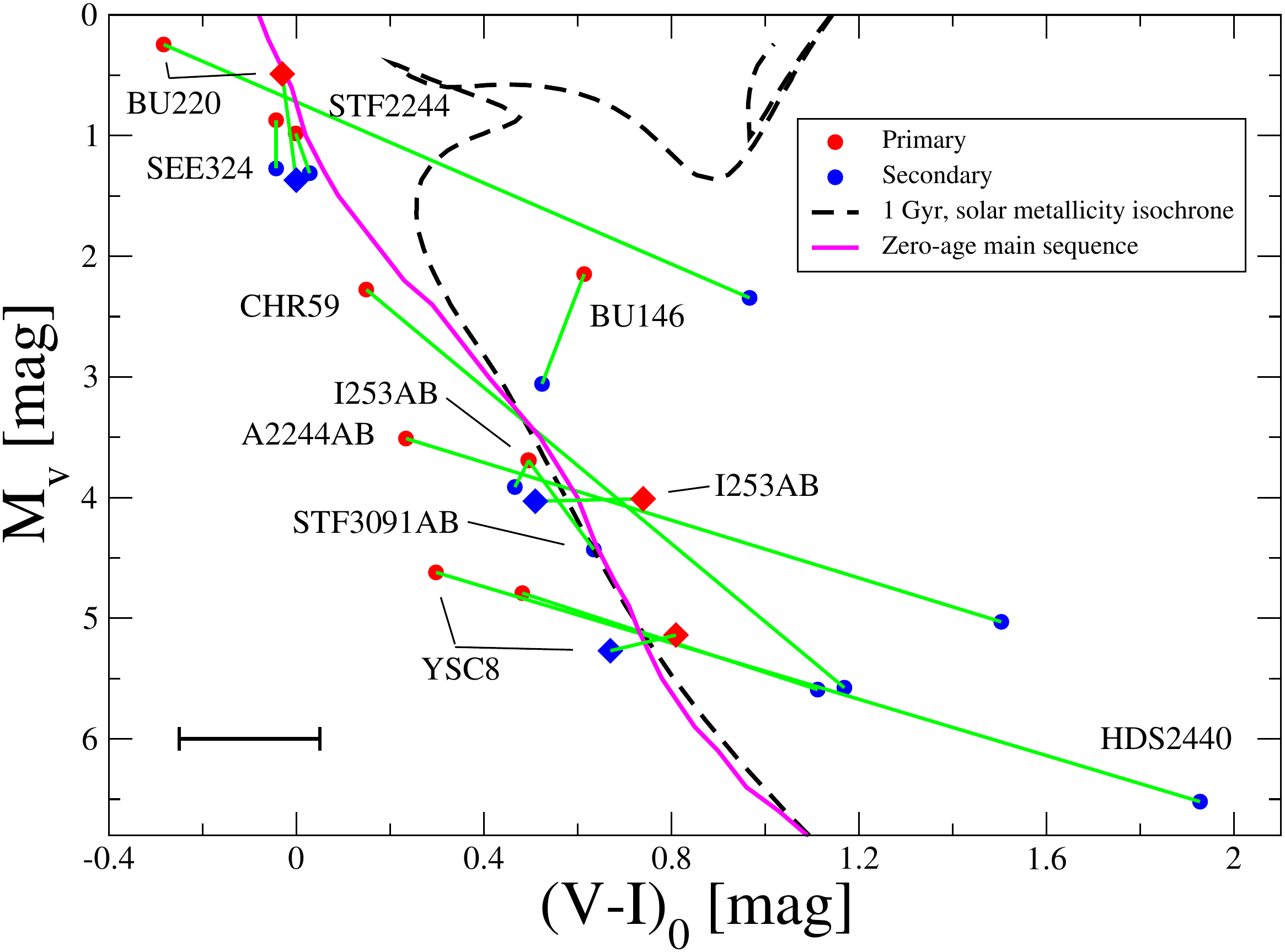}
\caption{H-R diagram for our sample of binary system. Each pair (red and blue circles) has been joined by a line, and the discovery designation is noted. The bar at (-0.1,+6.0) shows the estimated error of the photometry, as discussed in Section~\ref{sec:sample}. For YSC8, BU220, and I253AB we also show our "corrected" magnitudes (red and blue diamonds), obtained from our $\Delta y$ measurements as explained in Section~\ref{sec:sb2}. Clearly, these latter agree much better with the theoretical isochrone. See text for details and comments on individual systems.
\label{fig:hrdiag}}
\end{figure}

In Figure~\ref{fig:hrdiag} we present an observational H-R diagram for our visual binaries with available $V$ and $I$-band photometry, plus the spectroscopic binary YSC8. To derive $(V-I)$ colors for each component, we used the $V$ magnitudes for the primary and secondary given in Table~\ref{tab:photom1}, and for the $I$-band we proceeded as explained in Section~\ref{sec:comp}. As was explained in Section~\ref{sec:massum}, while interstellar extinction is small, it is not negligible from the point of view of placing our targets in an H-R diagram (particularly, for the colors). We have corrected each of our apparent magnitudes using the extinction in the $V$-band predicted by the \citet{MenVan1998} reddening model. Furthermore, the magnitudes in the $I$-band were corrected using the ratio $A_V/A_I= 3.1/1.50$ derived from Table~2 in that paper. Finally, to determine the absolute magnitudes, we adopted the published trigonometric parallaxes shown in Tables~\ref{tab:para} and~\ref{tab:dynplxYSC8} and our extinction-corrected apparent magnitudes. In the figure, red dots depict primary components and blue dots the corresponding secondaries. We note that at this scale the formal error in absolute magnitude, due to photometric uncertainties and the parallax, is negligible (of course, this does not consider possible systematic effects or biases on the parallaxes, which could be larger than the formal uncertainties). For reference, we  have also superimposed a 1~Gyr solar metallicity isochrone from \citet{Mariet2017}\footnote{available for download from \url{http://stev.oapd.inaf.it/cgi-bin/cmd}} (dashed black line) and an empirical  zero-age main sequence (ZAMS) from  \citet{Sch1982} (magenta solid line, kindly provided by G. Carraro\footnote{Personal communication}).

Based on Figure~\ref{fig:hrdiag} and our orbital fitting results, in what follows we present  comments on individual systems:

\begin{trivlist}
\item {\bf I292=WDS10043$-$2823:} The latest higher-precision data, mainly from HRCam, leads to a beautiful -but short- orbital arc. A broader orbital coverage is therefore still needed. Note that the Gaia measurement at epoch 2015.0 (which was given a weight of 1~mas) falls right on top of our computed orbit (see Figure~\ref{fig:examples} middle panel).\\

\item {\bf CVN16AaAb=WDS10174$-$5354:} This is a low-mass X-ray binary with a recent orbit from Tok2019c. We were able to improve the orbit thanks to recent observations with HRCam, at epochs 2018.24, 2018.98, and 2019.95. Our latest epoch is close to periastron passage, which is circa 2021. The individual component magnitudes had to be calculated through intermediate steps described in Section~\ref{sec:comp}. The rather large difference between the trigonometric and dynamical parallaxes, as well as between the dynamical and astrometric masses (see Table~\ref{tab:para}), could be due to uncertainties in the photometry of the individual components, since the orbit itself (Grade 2) is well determined. However, if instead of adopting a $V_{\mbox{Sys}}=13.92$ (see Section~\ref{sec:comp}), we adopt the faintest combined photometry allowed by the consulted surveys (i.e., $V_{\mbox{Sys}}= 14.16$ from ASAS in Table~\ref{tab:photom2}; see also Figure~\ref{fig:photom2} in Appendix~\ref{A1}), we obtain a mass sum of $0.33~M_\odot$ and a dynamical parallax of 46.26~mas, in marginally better agreement with our astrometric mass and the Gaia parallax, but still not comfortable. Finally, we cannot rule out an extant uncertainty in the trigonometric parallax for this object, since Hipparcos reported $57.0 \pm 0.7$~mas while Gaia DR2 reports $51.00 \pm 0.30$~mas (note however that adopting the Hipparcos parallax instead, would further aggravate the discrepancy between the dynamical and astrometric mass of the system).\\

\item {\bf BU220=WDS11125$-$1830:} We computed a relatively well-defined highly-inclined orbit. However, this is a long period system which lacks observations for about half of its orbit, (particularly when approaching apoastron, which will happen in about 150~yr - see upper panel of Figure~\ref{fig:examples}). We note the large discrepancy between the reported WDS magnitude difference ($V_{\mbox{S}}-V_{\mbox{P}}=2.12$) and our measured value, $\Delta y = 1.1$ (see Table~\ref{tab:photom1}), albeit there is a good coincidence between the dynamical and trigonometric parallaxes, and between the dynamical and orbital masses (see Table~\ref{tab:para}). If we keep the WDS system magnitude ($V_{\mbox{Sys}}=6.09$), but we instead adopt our measured value for $\Delta y$, we would expect that $V_{\mbox{P}}=6.49$ and $V_{\mbox{S}}=7.37$. In this case, the Dynamical mass increases to $4.47~M_\odot$ (as expected, since the secondary being brighter is now heavier), while the dynamical parallax becomes 6.35~mas, which is still within less than $1\sigma$ of the (Hipparcos) trigonometric parallax. Note that, with this corrected magnitudes, the location of both components on the H-R diagram coincides with the isochrone (see Figure~\ref{fig:hrdiag}).\\

\item {\bf FIN297AB=WDS13145$-$2417:} The orbit is well-constrained, thanks to a relevant coverage with HRCam near periastron at epochs 2017.53, 2018.24 and 2019.05. Actually, according to Table~\ref{tab:orbel1}, the most recent periastron passage occurred at $T_0 =2018.11$.\\

\item {\bf FIN370=WDS13574$-$6229:} A slight improvement in the formal error of the orbital elements (with respect to a recent orbit) was achieved, possibly due to our new HRCam data at epochs 2018.25 and 2019.05. The dynamical mass is probably somewhat erroneous, because the primary is a sub-giant. The parallax from Gaia DR2 ($8.06 \pm 0.92$~mas) could be biased. Hipparcos gives a parallax of $14.43 \pm 0.65$~mas for this system. Adoption of the Hipparcos parallax gives a more reasonable mass sum, of 2.4$M_\odot$, see Figure~\ref{fig:mass}.\\

\item {\bf STF3091AB=WDS15160$-$0454:} Highly inclined orbit. There is some ambiguity because of possible quadrant flips, albeit the magnitude difference between the components is not too small. The orbital mass is too large for its spectral type (F7V to F8V, with a secondary about G1V, for an expected mass sum of about 2.3~$M_\odot$.). A parallax from Gaia is not available. The Hipparcos parallax has a large error, but is coincident with the dynamical parallax within less than 2$\sigma$. An alternative least-squares orbit from Tokovinin\footnote{Personal communication.}, using basically the same data points, but a different choice of weights in particular for the older measurements, gives $P=214.44$~yr and $a=765$~mas, and a smaller eccentricity, but essentially the same angular parameters. That result leads to a more comfortable mass sum of 2.38~$M_\odot$.\\

\item {\bf A2176=WDS15420$+$0027:} Possible sub-giant (see Table~\ref{tab:photom1}) with a larger astrometric mass than it is implied by its luminosity. This is in agreement with our orbital solution and the dynamical mass (the latter assumes it is a main-sequence star). Note the good correspondence between the dynamical and trigonometric parallaxes, which reinforces this conjecture.\\

\item {\bf CHR59=WDS17005$+$0635:} This system is classified as Grade 4 on the USNO orbit catalogue. We find a large change in eccentricity from the previous orbit by \citet{RoMa2018}. This latter is based on adaptive optics observations from 2001 to 2006, and leads to $e= 0.017 \pm 0.021$ in contrast to our value of $0.0814 \pm 0.0070$.\\

\item {\bf YSC62=WDS17077$+$0722:} Our latest epoch from HRCam (2019.38) is very close to periastron (2020.6) and helped constrain the orbital parameters. Previous grade 3 classification can probably be upgraded to grade 2. Unfortunately the trigonometric parallax error is rather large; actually, the largest on our sample (see Table~\ref{tab:para}), albeit within less than 2$\sigma$ from the dynamical parallax. A Gaia parallax is not yet available. Note that the inter-quartile range for the mass sum (see last column of Table~\ref{tab:para}) is very tight, leading to a well constrained solution (see Figure~\ref{fig:examples} lower panel). In Figure~\ref{fig:hrdiagYSC62} we present a dedicated H-R diagram for this object. Its position on this diagram is discrepant with the 1 Gyr, solar metallicity isochrone. Considering that, according to SIMBAD, the metallicity for YSC62 could be as low as [Fe/H]$=-0.38$, in Figure~\ref{fig:hrdiagYSC62} we also include for reference a low-metallicity isochrone, but the discrepancy is similar. This object has a rather large proper motion of $(\mu_{\alpha},\mu_{\delta})=(-484.3 \pm 9.6,-385.4 \pm $9.5)~masy$^{-1}$, but given its small distance (only slightly beyond 12~pc) this translates into a modest transverse velocity of about 36~kms$^{-1}$, while the most recent reported radial velocity is 85~kms$^{-1}$. The total space velocity is indicative of old-disk or, at most, thick-disk kinematics, but not halo. In Section~\ref{sec:comp} we described the difficulties encountered in obtaining the photometry for the individual components for this system, which is subject to large uncertainties. We have consulted the "VizieR Photometry viewer"\footnote{Available at \url{http://vizier.unistra.fr/vizier/sed/}} to double check our system's photometry. This facility indicates combined fluxes for YSC62 of $1.04\times10^{-2}, 8.98\times10^{-3}$, and $7.40\times10^{-3}$~Jy from three different sources, while the combined $I$-band flux is $1.13\times10^{-1}$~Jy. Transforming these fluxes to Vega-mag apparent magnitudes using \citet{PicklesDepagne2010}, yields $V=13.86,14.02,14.23$~mag, and $I=10.87$. The mean $V$ mag is 14.04, equal to the system's value given by the WDS\footnote{as can be seen from Tables~\ref{tab:photom1}, \ref{tab:photom2} and the bottom panel on Figure~\ref{fig:photom3} in Appendix~\ref{A1}, the system's $V$-band photometry for this object seems well defined.}, but the color is however quite different from the one computed in Section~\ref{sec:comp} ($V-I=2.80$), ranging from 3.00 to 3.37 depending on the adopted $V$~mag. Of course, adopting a redder color will move the system to right in the H-R diagram, which is what we need judging from the isochrones. Indeed, if we adopt the reddest color, i.e., $V-I=3.37$, while maintaining the $V$-band photometry from WDS, we obtain the diamonds shown in the H-R diagram. The agreement is better, but still not comfortable. The only way to reconcile our photometry for the individual components with the isochrone would be if we adopt a large $\Delta I$ between primary and secondary. The solid squares in the figure show the expected location of both components for the same conditions as described above, plus those for a fictitious value of $\Delta I$ = 0.65, which renders a much better fit - this is meant only as an exercise to show that a plausible set of values can reach good agreement with the theoretical isochrones. One can also see that, if the overall system color $V-I$ were reliable, one could in principle discriminate between the two isochrones, which is not the case now. While it is known that measuring small magnitude differences at small separations is particularly difficult from Speckle photometry (leading to smaller than real magnitude differences), this proposed change in $\Delta I$ would imply a value larger by 3.5$\sigma$ than our measured value, which seems rather excessive. This example highlights how critical is to have not only good quality astrometry for orbits, but also reliable photometry for a complete astrophysical interpretation of a binary system.\\

\begin{figure}
\plotone{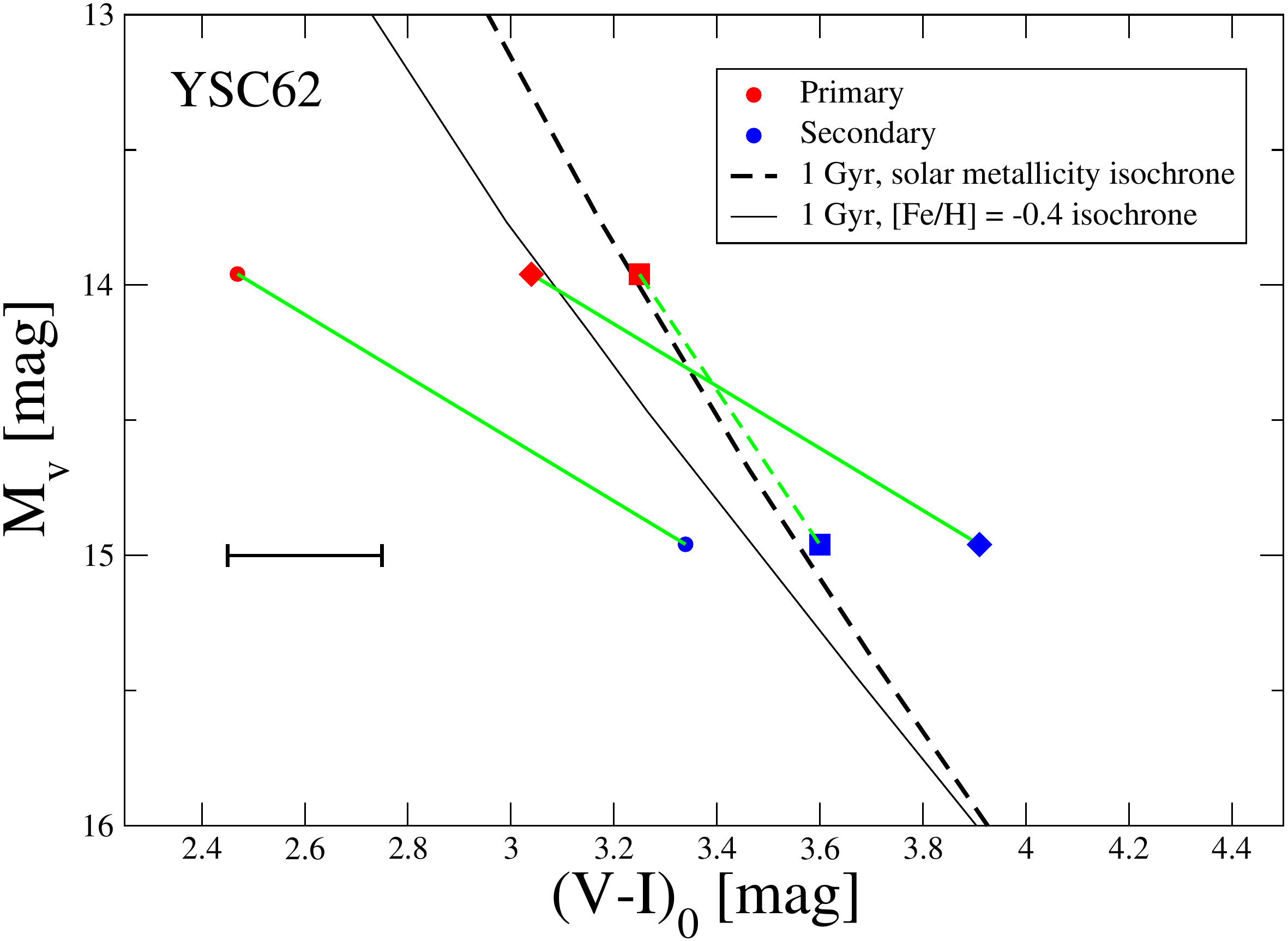}
\caption{H-R diagram for YSC62. Symbols are similar to those in Figure~\ref{fig:hrdiag}. The squares are for a fictitious value of $\Delta I = 0.65$ between primary and secondary, quite different from our measured value of $\Delta I = 0.13 \pm 0.15$. We also include a low metallicity isochrone, with [Fe/H]$=-0.4$.
\label{fig:hrdiagYSC62}}
\end{figure}

\item {\bf HDS2440=WDS17155$+$1052:} Nice orbit and good phase coverage. Our observations led to a small change of the orbital elements compared to \cite{Cveet2014}. Could be promoted to grade 2. The position of the primary and the secondary on the H-R diagram is somewhat uncomfortable. Unfortunately, unlike the case for BU220, I253AB, and YSC8 for which we could revise their photometry using our own measurements (leading to a much better agreement with the isochrones), we cannot do this for this target because we have not measured its $\Delta y$ (although we have a $\Delta I$ which allowed us to place it on the diagram). The system's magnitude seems well established (see Section~\ref{sec:photom} and the middle panel on Figure~\ref{fig:photom3}), and there is no indication that the object might be variable. Note however the rather large discrepancy between the more recent ASAS-SN magnitude for this object ($V=9.14 \pm 0.12$) in comparison to all other reported values ($V \sim 8.6-8.8$) - see Tables~\ref{tab:photom1} and~\ref{tab:photom2}, and the middle panel of Figure~\ref{fig:photom3} in Appendix~\ref{A1}. This photometric value should be viewed however with caution, since it is brighter than the nominal bright limit for ASAS-SN (10th mag). In fact, if one adopts this system's magnitude, the derived dynamical parallax and masses turn out to be 14.44~mas and 0.97$M_\odot$ and0.85$M_\odot$ respectively, leading to a larger discrepancy with the Gaia parallax and the orbital mass sum, as shown in Table~\ref{tab:para}).\\

\item {\bf SEE324=WDS17181$-$3810:} Our observations led to a large improvement in comparison to the previous preliminary (grade 5) orbit from \cite{Tokoet2015}, but still needs a more complete orbital coverage. The period seems better constrained in our new solution (it had $P=300$~yr, with no quoted error, in previous solution). Too large orbital mass for its spectral type (expected about 4$M_\odot$). Dynamical mass may be erroneous in the event it is a subgiant (see Table~\ref{tab:photom1}).\\

\item {\bf A2244AB=WDS17283$-$2058:} Our new orbit does not differ significantly from previously published orbit, but formal errors in all the orbital parameters were decreased significantly. The dynamical mass is quite larger than the orbital mass, probably because the dynamical parallax is more than 7$\sigma$ smaller than Gaia DR2 parallax, which could be biased since $a\sim 0.17$~arcsec, i.e., close to the resolution limit of Gaia. This interpretation however could be challenged given our photometric analysis in Section~\ref{sec:photom}, where we found that $V_{\mbox{APASS}}=8.875$, whereas our adopted system's magnitude for Table~\ref{tab:para} has $V_{\mbox{Sys}}=8.03$ (see Table~\ref{tab:photom1}). If we adopt the APASS value, the resulting dynamical parallax is 10.47~mas, with dynamical masses Mass$^{\mbox{\tiny{dyn}}}_{\mbox{\tiny{P}}}$=1.09$M_\odot$ and Mass$^{\mbox{\tiny{dyn}}}_{\mbox{\tiny{P}}}$=1.06$M_\odot$, much more consistent with the Gaia parallax of $10.68 \pm 0.11$~mas, and with the orbital mass sum of $2.03\pm0.04$~$M_\odot$ reported on Table~\ref{tab:para}.\\

\item {\bf STF2244=WDS17571$+$0004:} This is the system with the second longest period in our sample. Its mass seems too large for its spectral type and individual component apparent magnitudes, which imply a mass sum smaller than 3.7~$M_\odot$ (see Figure~\ref{fig:mass}). According to the USNO orbital catalogue, \citet{Malet2012} obtained dynamical, photometric, and spectroscopic masses of $8.83 \pm 5.28$, 4.60, and 2.00~$M_\odot$, respectively.  The latest value is the lower bound -if the secondary is very light- which is not in agreement with the WDS component magnitudes, nor with our measured $\Delta m$ (see Table~\ref{tab:photom1}). Note also the slightly more than $3\sigma$ discrepancy between its (Hipparcos) parallax, and the derived dynamical parallax in Table~\ref{tab:para} (see also Figures~\ref{fig:para}).\\

\item {\bf I253AB=WDS19190$-$3317:} Although our new observations led to a significant improvement in comparison to a previous orbit from \cite{vdBo1954}, we still consider our result as grade 4. The orbit is highly inclined (with a not so long period), which demands a better orbital coverage. Periastron is defined only by two micrometric observations at epochs 1939.59 and 1940.78. Interferometric observations on the next Periastron passage (circa 2025) are required. We note the large discrepancy between the reported WDS magnitude difference ($V_{\mbox{S}}-V_{\mbox{P}}=1.52$) and our measured value, $\Delta y = 0.20$ (see Table~\ref{tab:photom1}). This fact probably explains in part the large discrepancy (5$\sigma$ difference) between the derived dynamical parallax and mass, and the trigonometric parallax and orbital mass (see Table~\ref{tab:para}). The orbital mass is however too large for its spectral type (G1V); for small $\Delta y$ the system's mass should be close to 2$M_odot$, in agreement with the dynamical mass. In sum, the WDS individual component magnitudes seem indeed to be erroneous, producing a large dynamical parallax which compensates the bad photometry and leads to a reasonable dynamical mass simply by chance. An alternative least-squares orbit from Tokovinin\footnote{Personal communication.}, using a different choice of weights (specially for the older measurements) on basically the same data, gives $P=57.51$~yr and $a=419$~mas with a slightly smaller eccentricity, but essentially the same angular parameters. This leads to a reduced mass sum of 3.72~$M_\odot$, still too large, but less discrepant than ours. We finally note that our revised photometry lead to both components on the H-R diagram to coincides much better with the isochrone (see Figure~\ref{fig:hrdiag}).\\

\item {\bf BU146=WDS19471$-$1953:} Our observations cover a small orbital arc, so at this point more data is required to constrain the period. Unfortunately, observations previous to epoch 1940.0 have extremely large residuals, so they could not be used for the solution. The semi-major axis of the orbit is not so small ($\sim$ 1~arcsec), therefore the parallax from Gaia DR2 ($5.44 \pm 0.99$~mas) should not be strongly affected by orbital motion, albeit this value is quite different from the Hipparcos parallax of $10.36 \pm 0.91$~mas. In any case, the resulting mass sum is bad adopting either parallax. An alternative least-squares orbit from Tokovinin\footnote{Personal communication.}, using the parallax from Hipparcos and a different choice of weights (specially for the older measurements), leads to a mass sum of 3.21~$M_\odot$; still too large, but less discrepant than ours. The position on the H-R diagram might suggest that the primary is moving off the main-sequence, although it is listed as luminosity class V in SIMBAD. On the basis of photometric evidence, it is also classified as a main-sequence star by \citet{Eggen1986}. Note however that there does not seem to be any discrepancy between the magnitude difference reported by WDS and our measurements (see Table~\ref{tab:photom1}), therefore, if one is to believe our photometry, it is suggestive, looking at Figure~\ref{fig:hrdiag}, that an older isochrone might actually fit both points, the primary being on the sub-giant branch just above the turnoff. We finally remark the good consistency between the system's photometry for this source from different surveys, as shown in Tables~\ref{tab:photom1}, \ref{tab:photom2}, and the top panel on Figure~\ref{fig:photom3} in Appendix~\ref{A1}.

\end{trivlist}

\section{Conclusions} \label{sec:conc}

In this paper we present orbital elements and masses for fifteen visual binaries, and one SB2 system, based on observations from our on-going HRCam@SOAR Speckle survey, supplemented with historical data from the Washington Double Star catalogue maintained by the US Naval Observatory. For one object, I292, we have also incorporated a Gaia astrometric measurement which, at a separation of 0.75~arcsec, is found to be completely consistent with our measurements at two nearby epochs that bracket the Gaia epoch.

The orbits have been computed using our MCMC code, duly described in \citet{Mendezetal2017} and \citet{Clavet2019}. The quality of our final orbital elements is variable, ranging from tentative to good quality, depending on orbital coverage and overall data quality. Their periods span a range from slightly more than 5~yr to more than 500~yr; and their spectral types go from early A to mid M -implying system masses from slightly more than 4~$M_\odot$ down to 0.2~$M_\odot$. They are located at distances between about 12 and less than 200~pc, mostly at low Galactic latitude.

We also present the first combined orbit, individual component masses, and orbital parallax for the SB2 system YSC8 derived from a dynamically self-consistent MCMC fit to the joint astrometric and radial velocity data available. We obtained individual masses of $0.897 \pm 0.027~M_\odot$, and $0.857 \pm 0.026~M_\odot$, and an orbital parallax of $26.61 \pm 0.29$~mas, which compares very well with Gaia DR2 trigonometric parallax of $26.55 \pm 0.27$~mas. This proves, once again, that SB2 binaries with astrometric orbits are exquisite astrophysical laboratories, because they allow us to characterize the system independently of its trigonometric parallax, as discussed in a very comprehensive study of SB2 systems with astrometry by \citet{Picco2020} (see in particular, their Figure~2).

We have done a thorough comparison of the photometry for our targets, using data from several photometric surveys to asses the quality of our system's combined photometry. In combination with precise trigonometric parallaxes from Gaia, we use this photometry to address their evolutionary status by placing our objects in an H-R diagram which highlights the paramount importance of good quality photometry in terms of not only good magnitude differences (through Speckle or other techniques), abut also the system's apparent magnitudes and colors.

As stated in Section~\ref{sec:intro}, one of the long term goals of our Speckle program is to provide empirical points for more detailed investigations of the MLR. The most reliable MLR, based on the best measured binary systems, is that of \citet{Torreset2010}. Figure~6 on that paper (which encompasses the mass range of our binaries) shows a significant scatter in luminosity at each mass value, which is customarily assumed to be due to the combined effects of abundance differences and stellar evolution. Typical error bars on that plot are 5\% in mass (see also their Table~2), and 0.05 in $\log L/L_\odot$, which seems clearly sufficient to explore the intrinsic scatter on the MLR. Assuming we are limited by random (and not systematic) errors, simple error propagation on both Kepler's law and the distance-modulus equation allows us to estimate the required uncertainties on the relevant observational parameters to obtain a pre-specified final uncertainty in mass and luminosity. For this calculation we have assumed a typical parallax uncertainty of 1\% for our targets (0.04~mas for $G \le 14$ according to \citet{Luriet2018}, and a distance limit of 250~pc for the targets on our sample)\footnote{Note however that, according to Table~\ref{tab:para}, some objects, specially those without Gaia DR2 parallaxes, have uncertainties much larger than 1\%.}. In this case, a 1\% uncertainty in both the semi-major axis and the period renders an uncertainty of about 5\% in (total) mass, i.e., similar to that of \citet{Torreset2010}. A quick look at Table~\ref{tab:orbel1} shows that several objects do satisfy these criteria. Of course, this can also be corroborated directly on our final derived masses, given on Table~\ref{tab:para}, which already fully account for the overall uncertainties of all the relevant parameters, including parallax. In the case of the SB2 YSC8, the mass uncertainty of the individual components is 3\% (Table~\ref{tab:dynplxYSC8}). On the other hand, error propagation shows that in order to achieve an uncertainty of 0.05 in $\log L/L_\odot$, the uncertainty on the magnitude of the individual components (and the inter-stellar extinction) must be smaller than $\sim$0.03~mag. Unfortunately, in this case, we are in a much worse situation, basically due to the poorly measured magnitude differences (typical uncertainties of $\sim$0.1~mag ), as extensively discussed in Section~\ref{sec:comp}, leading to uncertainties three times larger than desired in $\log L/L_\odot$. Therefore, as already emphasized, a much better effort should be spent refining the individual component magnitudes of the binary systems with well defined masses.

Visual binary research is one of the oldest branch of observational quantitative astronomy, dating back at least to $\sim$1650, when the Italian astronomer Giovanni Riccioli discovered the first visual binary system, Mizar, in Ursa Major \citep{Niem2001}. From its very beginnings, this branch of astronomy has required measurements over time to fulfil its goals, and therefore it could be considered the precursor of the currently fashionable “time-domain astronomy”, preceding it by more than three centuries. It is also an inter-generational science, because determining the orbits of long period systems (sometimes as long as several centuries) requires the continuous effort of many generations of unselfish observational astronomers. In this tradition, continued monitoring of binary systems with HRCam@SOAR, is contributing to our knowledge of orbits and masses of binary stars, and aiding in our quest to understand the effects of age and metallicity on the MLR.

\section{Acknowledgements}

RAM and EC acknowledge support from CONICYT/FONDECYT Grant Nr. 1190038 and from the Chilean Centro de Excelencia en Astrof\ia sica y Tecnolog\ia as Afines (CATA) BASAL PFB/06. We are indebted to Drs. Andrei Tokovinin (Cerro Tololo Inter-American Observatory) and Elliott Horch (Southern Connecticut State University) for their continued support of this program, and to an anonymous referee who provided numerous suggestions that have significantly improved the readability of the paper, and that also lead to the incorporation of Appendix~\ref{A1}.

This research has made use of the Washington Double Star Catalog maintained at the U.S. Naval Observatory and of the SIMBAD database, operated at CDS, Strasbourg, France. This research was made possible through the use of the AAVSO Photometric All-Sky Survey (APASS), funded by the Robert Martin Ayers Sciences Fund and NSF AST-1412587. This work has made use of data from the European Space Agency (ESA) mission Gaia (\url{https://www.cosmos.esa.int/gaia}), processed by the Gaia Data Processing and Analysis Consortium (DPAC, \url{https://www.cosmos.esa.int/web/gaia/dpac/consortium}). Funding for the DPAC has been provided by national institutions, in particular the institutions participating in the Gaia Multilateral Agreement. We are very grateful for the continuous support of the Chilean National Time Allocation Committee under programs CN2018A-1, CN2019A-2, CN2019B-13, and CN2020A-19.

\appendix
\section{Photometric consistency} \label{A1}

\begin{figure}[ht!]
\plotone{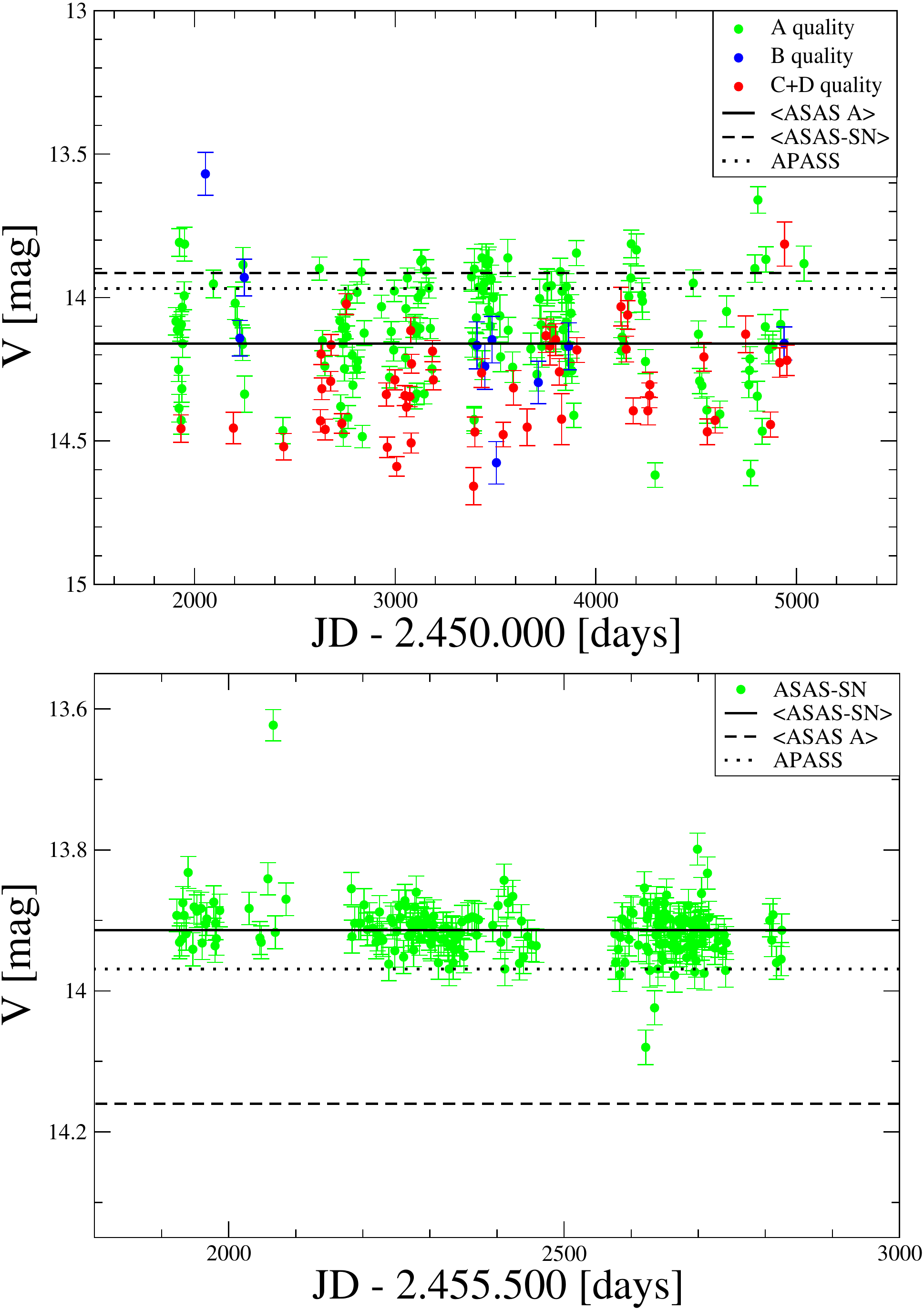}
\caption{$V$-band light curves for CVN16AaAb from ASAS (top panel) and ASAS-SN (lower panel) photometry, with their reported uncertainties. In both plots the solid line depicts the mean of all plotted values. In the top plot, the dashed line is the mean of the ASAS-SN data, while the dotted line is the mean value of the APASS data (this latter is also shown on the lower panel). In the lower panel however, the dashed line is the mean value of the ASAS A-quality data. The rms of the ASAS-SN data is clearly much smaller than that of the ASAS survey. Note the different range for the ordinate of the  plots, and also that these surveys were not contemporary (different JD zero points). As it can be seen from Table~\ref{tab:photom1} the photometry given in SIMBAD is basically the same as that reported by ASAS, and therefore it was not included in the plot. Additionally, the synthetic $V$ photometry from PD2001 presented in Table~\ref{tab:photom2} is completely wrong, causing it to lie off the scale of these plots (the same is true in the $I$-band; compare $i'$ to $I_{\mbox{PD2010}}$ in that table, see also Section~\ref{sec:photom}).}\label{fig:photom2}
\end{figure}

As can be seen from Table~\ref{tab:photom2}, the only object for which we could reliably compare simultaneously the photometry from ASAS, ASAS-SN and APASS, is CVN16AaAb, which is the second faintest object in our sample. The light curves from ASAS and ASAS-SN for this object are shown in Figure~\ref{fig:photom2}. The ASAS catalogue classifies the individual photometric points into four categories, A, B, C and D, depending on the quality of the night (from best to worst). In the figure (top panel) we have split these data using different colors, and it is apparent that the rms of the A-quality data is smaller than that of the C+D photometry. However, the mean photometric values (not shown) are indistinguishable, within their errors, from the mean for the A-quality data ($<V_{\mbox{ASAS}}>_A = 14.11 \pm 0.18$, shown as a solid line in the figure), and this value is in turn indistinguishable from the mean for the whole data set, presented in Table~\ref{tab:photom2}. A comparison of the data and their errors on the upper and lower panels, shows a difference of almost 0.25~mag (see also Table~\ref{tab:photom2}) between ASAS and ASAS-S, which, while clearly visible in the plots, is only 1.4$\sigma$ of the ASAS rms and therefore not significant. The APASS value lies between the two values, being closer to the ASAS-SN photometry. Furthermore, from the light curves, and the consistency between all the reported values, there is no indication that CVN16AaAb might be variable. An analysis of the possible impact of these different photometric values on the calculated dynamical parallax and masses for this system is presented in Section~\ref{sec:hr}).

\begin{figure}[ht!]
\plotone{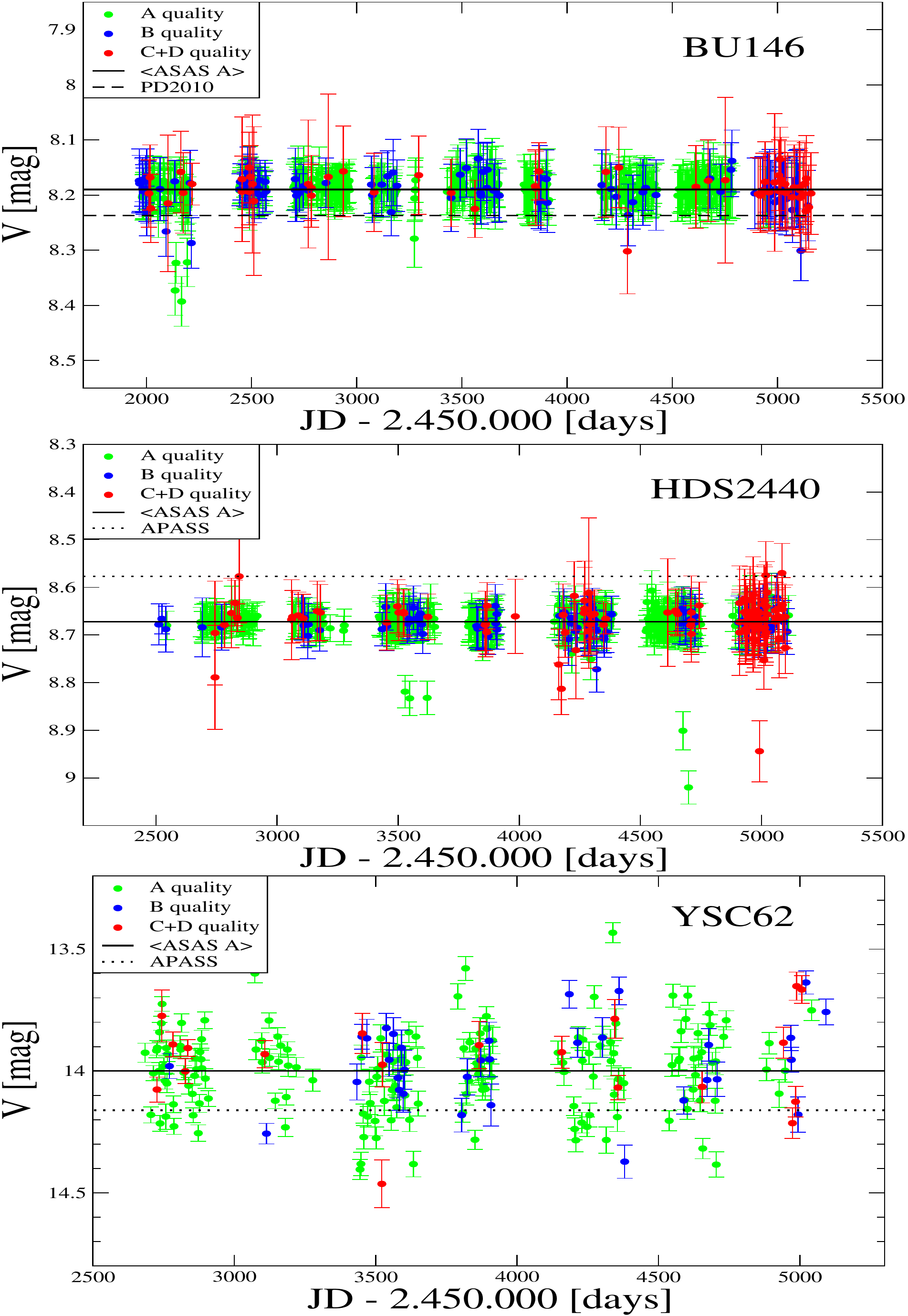}
\caption{$V$-band lights curves from ASAS for BU146 (top panel), HDS2440 (mid panel) and YSC62 (bottom panel). Different quality data from ASAS is depicted with different colors, as indicated. The solid line is the average $V$-band photometry from the A-quality data. Values from Hipparcos, SIMBAD, APASS, or PD2010 that lie too close to the average were omitted for clarity; only values different from the average of the ASAS A-quality data are plotted, as follows: APASS values with a dotted line and PD2010 with a dashed line (top panel).\label{fig:photom3}}
\end{figure}

Besides CVN16AaAb, Table~\ref{tab:photom2} shows that the only remaining targets for which we can compare ASAS and APASS with confidence are BU146, HDS2440, and YSC62 (the first two are above the bright magnitude limit for ASAS-SN, while the last one is not in this photometric database). This comparison is shown in Figure~\ref{fig:photom3}. The maximum differences in the mean $V$~mag for BU146, HDS2440, and YSC62 among the consulted surveys are 2.2$\sigma$, 3.3$\sigma$ and 1$\sigma$, where 1$\sigma$ is the largest uncertainty value reported (or calculated) from the involved surveys for each dataset, and given in Table~\ref{tab:photom2}.

\bibliography{Orbits}{}
\bibliographystyle{aasjournal}

\newpage

\facilities{CTIO: SOAR 4.0m}
\software{IRAF~\citep{Tody2000}, TopCat~\citep{Taylor2020}, Grace~\citep{Stambu}}

\end{document}